\documentclass{IEEEtran}
\usepackage[dvipsnames]{xcolor}
\usepackage{cite}
\usepackage{amsmath,amssymb,amsfonts}
\usepackage{mathtools}
\usepackage{lipsum}
\usepackage{algorithmic}
\usepackage{graphicx}
\usepackage{textcomp}
\usepackage{booktabs}
\usepackage{esint}
\usepackage{cuted}
\def\BibTeX{{\rm B\kern-.05em{\sc i\kern-.025em b}\kern-.08em
    T\kern-.1667em\lower.7ex\hbox{E}\kern-.125emX}}
\begin{document}
\title{Semi-analytical Model of Multi-tile Rectangular Waveguide-fed Metasurfaces using Coupled Dipole Modeling Framework}
\author{Insang Yoo, \IEEEmembership{Member, IEEE}, Michael Boyarsky, \IEEEmembership{Member, IEEE}, and David R. Smith \IEEEmembership{Senior Member, IEEE}
\thanks{I. Yoo is with the School of Electrical and Electronic Engineering, Yonsei University, Seoul, Korea. He was previously with the Department of Electrical and Computer Engineering and the Center for Metamaterials and Integrated Plasmonics, Duke University, Durham, NC, 27708 USA (e-mail: insang.yoo@yonsei.ac.kr). M. Boyarsky and D. R. Smith are with the Department of Electrical and Computer Engineering and the Center for Metamaterials and Integrated Plasmonics, Duke University, Durham, NC, 27708 USA.}}

\maketitle

\begin{abstract}
We present a semi-analytical model to analyze multi-tile metasurface antennas consisting of a set of metasurface tiles and a practical power-dividing network that excites the tiles. The metasurface tiles consist of arrays of rectangular waveguides with subwavelength metamaterial radiators etched into their top walls, each of which can be accurately modeled as polarizable dipoles. The feed structure for the arrays comprises a slotted waveguide attached to their bottom wall, with coupling slots inserted into the common wall that are likewise modeled as polarizable dipoles. The proposed semi-analytical model employs a coupled-dipole framework that accurately captures dipolar interactions among constituent elements within the metasurface tiles, along with a multi-port network analysis technique that accounts for electromagnetic interactions between the tiles and the power divider, thereby forming a self-consistent formulation. The proposed model enables the prediction of key performance metrics, including overall S-parameters, radiation patterns, and gain, and is validated through full-wave numerical simulations. By significantly reducing the computational complexity associated with electrically large apertures, the proposed framework enables rapid and efficient modeling of the overall structure, thereby facilitating iterative optimization. The proposed model has potential applications as an efficient forward model for the design of wireless systems requiring large-aperture metasurface antennas, including remote sensing and next-generation wireless communication networks.
\end{abstract}


\begin{IEEEkeywords}
Aperture antenna, Leaky-wave antenna, Metasurfaces, Metamaterials.
\end{IEEEkeywords}

\section{Introduction} \label{sec:introduction}


\IEEEPARstart{M}{}etasurfaces have proven to be a promising physical platform for controlling electromagnetic radiation through structurally simple, lightweight, and low-cost architectures \cite{smith2017analysis,Maci2007,Fong2010,maci2011metasurfing,Patel2013,Quarfoth2013,Patel2013a,minatti2015modulated,epstein2016cavity,epstein2017arbitrary,lee2019method,budhu2020perfectly}. Among various configurations, waveguide-fed metasurfaces have attracted significant interest as an efficient platform for realizing electrically large, array-like antenna systems. These structures employ waveguide- or cavity-backed architectures loaded with subwavelength metamaterial radiators excited directly by guided modes, thereby eliminating the need for complex feed networks, costly RF components (e.g., phase shifters), and the associated calibration procedures required in conventional phased-array systems. Such advantages become increasingly significant as the aperture size of the antenna system increases. Consequently, waveguide-fed metasurfaces are highly attractive for a variety of modern wireless applications, including computational imaging \cite{hunt2013metamaterial}, wireless power transfer \cite{smith2017wpt}, and communication systems \cite{johnson2015sidelobe,yoo2018enhancing}. More recently, there has been growing interest in waveguide-fed metasurfaces with dynamic radiation-control capabilities, referred to as dynamic metasurface antennas (DMAs), which enable functionalities such as beam steering and wavefront shaping for applications including remote sensing and synthetic aperture radar systems \cite{boyarsky2017synthetic}.



Leveraging these advantages, waveguide-fed metasurfaces have recently been employed as building blocks for electrically large aperture systems, aligning closely with recent efforts to enhance the performance of wireless applications through the use of large-scale array antennas---including human-scale imaging systems \cite{gollub2017large,imani2020review}, holographic massive or extremely large-scale MIMO systems \cite{shlezinger2021dynamic,yoo2023sub,carlson2024hierarchical,castellanos2025embracing,gavriilidis2025near}, and synthetic aperture radar systems \cite{boyarsky2017synthetic}, among others. A straightforward yet effective approach to realizing such large apertures is to arrange waveguide-fed metasurfaces of appropriate dimensions in an array to form a composite aperture, as demonstrated in \cite{gollub2017large}. This array, or multi-tile configuration, not only enables modular design, fabrication, and troubleshooting of individual metasurface tiles---thereby improving the overall development and integration process---but also provides additional degrees of freedom, such as tile orientation and placement, for achieving desired aperture functionalities. Accordingly, the design of multi-tile metasurfaces is of significant interest for next-generation wireless systems using them.

\begin{figure}[!t]
\centering
\includegraphics[width=3.2in]{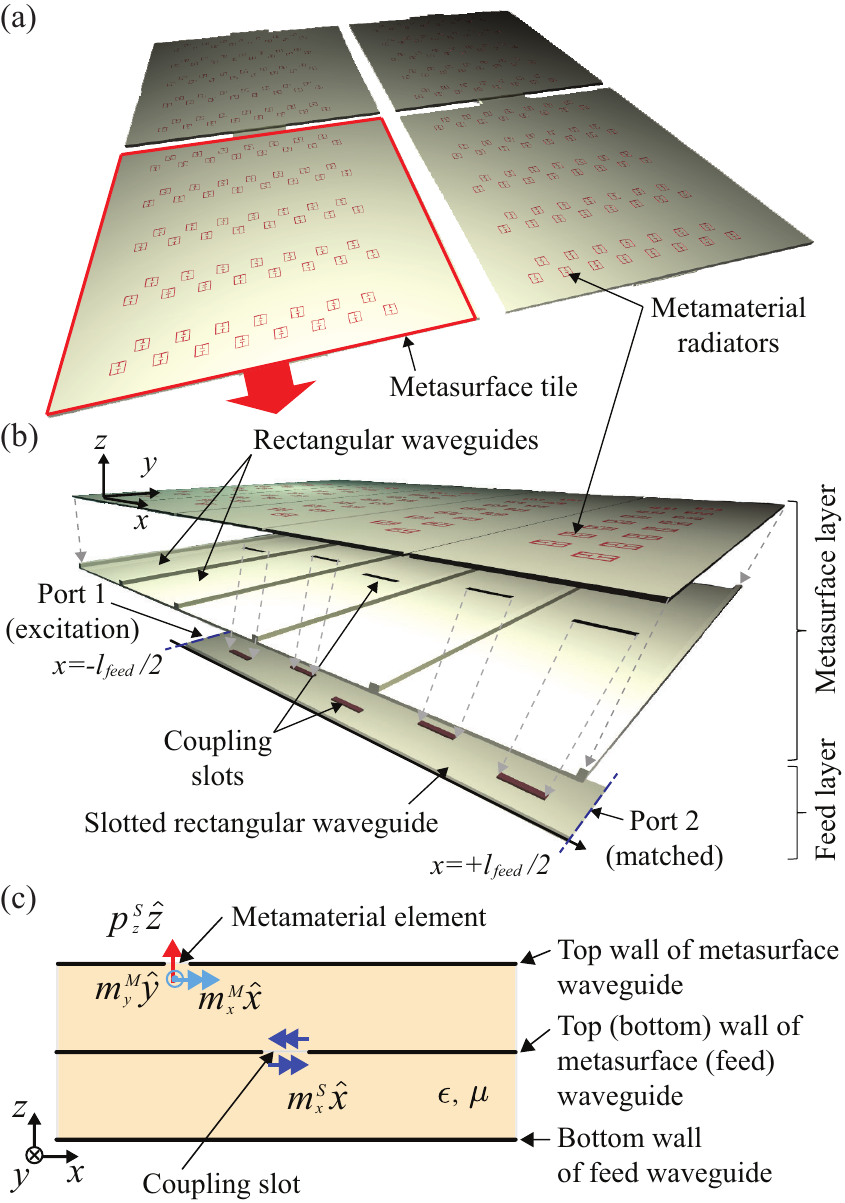}
\caption{(a) Example schematic of a multi-tile metasurface system. Each metasurface tile consists of multiple rectangular waveguide-fed metasurfaces, with subwavelength-sized metamaterial radiators etched into the top plates of the waveguides. (b) Exploded view of a single-tile metasurface. The rectangular waveguide metasurfaces are excited via coupling slots inserted into the common walls between the metasurfaces and a slotted feed waveguide. Port 1 of the feed waveguide serves as the input port, while port 2 of the feed waveguide and both ends of the metasurface waveguides are terminated with matched loads to suppress reflections. (c) Side-view schematic illustrating the dipolar modeling of metamaterial elements and coupling slots. An individual metamaterial radiator is modeled using in-plane magnetic dipoles and an out-of-plane electric dipole, and the dipoles acting above the aperture plane are omitted for brevity. The magnetic dipoles representing a coupling slot above and below its aperture plane have equal amplitudes and opposite directions.}
\label{Fig1_Schematic}
\end{figure}

In the context of beamforming and beam steering, arrayed rectangular waveguide– or equivalently, substrate-integrated waveguide (SIW)-fed metasurfaces have demonstrated their effectiveness as modular tiles for achieving wide scan angles and high aperture efficiency \cite{boyarsky2021electronically,boyarsky2023cascaded}. For instance, the metasurface antenna reported in \cite{boyarsky2023cascaded} employs an array of rectangular waveguide-fed metasurfaces excited by a slotted waveguide attached to the bottom wall of the metasurfaces. In this configuration, the slotted waveguide is used to deliver the input RF power to the upper metasurfaces through coupling slots inserted into the common walls between the metasurfaces and the feed waveguide (see Fig. \ref{Fig1_Schematic}). By appropriately adjusting the geometry and/or orientation of the coupling slots, efficient power coupling to the metasurfaces can be achieved, along with phase diversity of the fields exciting the metasurfaces---an effective method for suppressing unwanted grating lobes \cite{boyarsky2020grating}. Furthermore, the coupling slots are located near the center of each metasurface, as illustrated in Fig. \ref{Fig1_Schematic}(b), and the RF power injected through each slot is divided to propagates bidirectionally along the waveguide forming the metasurface. The bidirectional power distribution reduces the local power incident on individual metamaterial radiators, thereby improving the overall power-handling capability. Such improvement in power handling is particularly important for dynamically tunable metasurfaces incorporating switchable components (e.g., varactor or PIN diodes).


Accordingly, a multi-tile metasurface configuration---each tile comprising an array of rectangular waveguide–backed metasurfaces excited by a slotted feed waveguide---offers significant hardware advantages and is well suited for realizing electrically large antenna systems with beamforming and steering capabilities. Consequently, the development of an accurate electromagnetic model and an efficient design methodology for this configuration is crucial. The need is particularly pronounced because arrayed metasurface tiles are electrically large and include many subwavelength metamaterial radiating elements, thereby rendering full-wave analysis computationally intensive in both memory and runtime. To mitigate these challenges, an electromagnetic model, referred to as the cascaded network model, has recently been proposed \cite{boyarsky2023cascaded}. In this approach, the metasurface is segmented into short rectangular waveguide sections, each containing an embedded metamaterial radiator whose effective polarizability determines the associated network parameters (e.g., ABCD- or T-parameters); this equivalence is verified in \cite{smith2024equivalence}. By cascading these network representations, the overall electromagnetic response of the metasurface---including mutual coupling between metamaterial radiators \textit{only} through guided modes---can be efficiently analyzed.

While effective, the model presented in \cite{boyarsky2023cascaded} does not fully capture the electromagnetic interactions among all pairs of metamaterial elements within and across metasurface tiles, particularly those through the radiated fields. In modeling multi-tile metasurface systems fed by practical power-dividing networks, the mutual coupling between metamaterial elements should be considered for accurate prediction of the overall S-parameters. Specifically, to compute the overall S-parameters of the multi-tile system using the method in \cite{boyarsky2023cascaded}, it is required to characterize the metasurface tiles---without its power dividing feeder---as a multi-port network and then cascade it with the power divider. However, neglecting the mutual coupling between metamaterial radiators introduces errors in the S-parameter of the multi-port network representing the metasurface tiles, which subsequently propagate through the cascading process. These errors become increasingly significant as the aperture size of the multi-tile system increases, potentially leading to substantial inaccuracies in the analysis. Accordingly, an accurate and systematic treatment of the electromagnetic interactions is essential for the design and analysis of multi-tile metasurface configurations, particularly considering that constituent elements---including metamaterial radiators and coupling slots---are spaced at subwavelength distances.

This paper aims to address this gap by introducing a semi-analytical model---using a coupled-dipole framework---for multi-tile waveguide-fed metasurface systems excited by a practical power divider. Within this framework, the metamaterial elements and coupling slots are assumed to be electrically small, and their scattering behaviors are modeled as those of polarizable point dipoles, with mutual interactions accounted for via appropriate Green’s functions. The proposed coupled-dipole model for multi-tile metasurface systems captures \textit{all} possible dipolar interactions among the constituent elements forming the multi-tile system. Additionally, a multi-port network analysis is employed and combined with the dipole model to account for electromagnetic coupling between the metasurface tiles and the feeding power divider, thereby enabling the prediction of key performance metrics, such as overall S-parameters, radiation patterns, and gain of the multi-tile system. The proposed model is validated through full-wave analyses of design examples, with dimensions selected based on computational cost and resource constraints. Consequently, we suggest that the proposed model can be combined with numerical optimization techniques to establish a systematic design methodology for the multi-tile metasurface configuration.

\section{Semi-Analytic Model of Multi-tile Metasurface System} \label{sec:thoery}

\subsection{Coupled Dipole Model of a Single-tile Metasurface} \label{cdm_model}

Figure \ref{Fig1_Schematic}(a) illustrates an example schematic of a multi-tile metasurface antenna system, in which each metasurface tile consists of an array of rectangular waveguide-fed metasurfaces and a slotted rectangular waveguide. Figure \ref{Fig1_Schematic}(b) shows the configuration of a single metasurface tile used to form the multi-tile metasurface system. As shown in Fig. \ref{Fig1_Schematic}(b), the slotted rectangular waveguide is attached to the bottom wall of the metasurface array and serves as the feed waveguide. When port 1 is excited, the injected RF power is coupled through slots inserted into the common wall between the metasurface array and the feed waveguide. The coupled signal then propagates along the waveguides (i.e., in the $\pm\hat{y}$ directions) forming the metasurface and radiates into free space via metamaterial radiators. Both ends of the waveguides used to form metasurfaces are assumed to be terminated with matched loads to suppress reflections. Note that, although not explicitly shown for generality, the metasurface tiles depicted in Fig. \ref{Fig1_Schematic}(a) are assumed to be fed by a practical power-dividing network, whose output ports are connected to port 1 of each metasurface tile.

\begin{figure*}[!t]
\centering
\begin{equation} \label{DDA_eq_singletile}
\begin{aligned}
\LaTeXunderbrace{\begin{bmatrix*}[l]
\mathbf{G}^{mm,MM}_{xx} & \mathbf{G}^{mm,MM}_{xy} & \mathbf{G}^{me,MM}_{xz} & -\mathbf{G}^{mm,MS}_{xx} \\ 
\mathbf{G}^{mm,MM}_{yx} & \mathbf{G}^{mm,MM}_{yy} & \mathbf{G}^{me,MM}_{yz} & -\mathbf{G}^{mm,MS}_{yx} \\ 
\mathbf{G}^{em,MM}_{zx} & \mathbf{G}^{em,MM}_{zy} & \mathbf{G}^{ee,MM}_{zz} & -\mathbf{G}^{mm,MS}_{zx} \\ 
-\mathbf{G}^{mm,SM}_{xx} & -\mathbf{G}^{me,SM}_{xy} & -\mathbf{G}^{me,SM}_{xz} & \mathbf{G}^{mm,SS}_{xx}
\end{bmatrix*}}_{\triangleq\mathbf{G}_{ms}}
\LaTeXunderbrace{\begin{bmatrix*}[l]
\mathbf{m}_{x}^{M} \\
\mathbf{m}_{y}^{M} \\
\mathbf{p}_{z}^{M} \\
\mathbf{m}_{x}^{S}
\end{bmatrix*}}_{\triangleq\mathbf{J}_{ms}}
=\LaTeXunderbrace{\begin{bmatrix*}[l]
\mathbf{0} \\
\mathbf{0} \\
\mathbf{0} \\
\mathbf{H}^{inc}_{x}
\end{bmatrix*}}_{\triangleq\mathbf{F}_{ms}},
\end{aligned}
\end{equation}
\vspace*{-0.25in}
\end{figure*}


To develop a semi-analytical framework for the multi-tile metasurface antenna, we begin by introducing several assumptions regarding the geometry of the constituent elements. First, the dimensions of the rectangular waveguides and the dielectric material filling them are selected such that the dominant TE$_{10}$ mode is the only propagating mode within the operating band. Second, the metamaterial elements and coupling slots are assumed to be electrically small, so that their scattering behavior is dominated by dipolar responses \cite{collin1960field,bethe1944theory}. Under these assumptions, the scattering from individual elements can be well approximated as that by a combination of polarizable point electric and magnetic dipoles, each characterized by effective polarizability \cite{pulido2017polarizability,pulido2018analytical}. For example, as illustrated in Figs.~\ref{Fig1_Schematic}(b) and (c), the coupling slots can be modeled predominantly as magnetic dipoles oriented along their principal axis (i.e., the $\hat{x}$-direction), provided that their axial ratio is sufficiently large \cite{collin1960field}. Similarly, the metamaterial elements etched into the top wall of the waveguides---thereby forming the metasurfaces---can, in general, be represented by magnetic dipoles oriented along the $\hat{x}$ and $\hat{y}$-directions, as well as an electric dipole oriented along the $\hat{z}$-direction \cite{bethe1944theory,pulido2017polarizability,pulido2018analytical} (see Fig.~\ref{Fig1_Schematic} (c)). By appropriately tailoring the element geometry, one or more of these dipole components can be suppressed, thus making their contributions to the scattered fields negligibly small \cite{yoo2022experimental}. Note that the coupled-dipole framework developed in this work originates from the pioneering studies on the discrete dipole approximation (DDA) \cite{draine1988discrete}, which was subsequently adapted for the modeling of metamaterial devices \cite{bowen2012using} and waveguide-backed metasurfaces \cite{pulido2017polarizability,pulido2018analytical}. However, it should be emphasized that the present formulation for the dipole model extends these approaches by incorporating a dipolar treatment of coupling slots, thereby enabling the semi-analytical modeling of multilayered metasurfaces, illustrated in Fig. \ref{Fig1_Schematic}(b), which requires the modeling of coupled-waveguide structures. The coupled-waveguide structures have been extensively investigated in the design of power-dividing networks, filters, and couplers \cite{pozar2009microwave,collin1960field}, where the analyses typically rely on the circuit models or \textit{static} polarizabilities of irises embedded in waveguide walls, in contrast to the effective dynamic polarizabilities employed in this work.

Under these assumptions, a coupled dipole framework can be employed to develop a semi-analytical model of a single-tile metasurface comprising $N_{S}$ rectangular waveguides, each containing $N_{M}$ metamaterial radiators. Specifically, the dipole model can be formulated as a system of coupled equations, as given in (\ref{DDA_eq_singletile}). In (\ref{DDA_eq_singletile}), $\mathbf{m}_{x}^{M}$, $\mathbf{m}_{y}^{M}$ and $\mathbf{p}_{z}^{M}\in\mathbb{C}^{N_{M}N_{S} \times 2}$ represent the effective magnetic and electric dipole moments representing the metamaterial elements, respectively. Also, $\mathbf{m}_{x}^{S}\in\mathbb{C}^{N_{S} \times 2}$ denotes the magnetic dipole moments modeling the coupling slots in the feed waveguide. The superscripts $``M"$ and $``S"$ correspond to the metamaterial elements and slots, respectively. In (\ref{DDA_eq_singletile}), the entries of $\mathbf{H}^{inc}_{x}\in\mathbb{C}^{N_{S} \times 2}$ represent the magnetic fields incident at the center of the coupling slots, through which the metasurfaces and metamaterial elements are excited. Note that the entries of the matrix on the right-hand side are zero, except for the incident magnetic fields $\mathbf{H}^{inc}_{x}$. This is because the metamaterial radiators are excited by propagating waves within the waveguides of the metasurfaces, which are launched through the coupling slots.

Additionally, the off-diagonal entries of the $\mathbf{G}^{MM}$ matrices in (\ref{DDA_eq_singletile}) are the sum of the free-space Green’s functions and those of the rectangular waveguide, as proposed in \cite{pulido2017polarizability,yoo2022conformal}. In contrast, the off-diagonal elements of the $\mathbf{G}^{MS}$, $\mathbf{G}^{SM}$ and $\mathbf{G}^{SS}$ matrices include only the Green’s functions within the rectangular waveguide, as there is no mechanism for mutual coupling between the dipoles via radiated fields. Also, note that the diagonal elements of the matrices $\mathbf{G}^{mm,MM}_{xx}$, $\mathbf{G}^{ee,MM}_{zz}\in\mathbb{C}^{N_{M}\times N_{M}}$, and $\mathbf{G}^{mm,SS}_{xx}\in\mathbb{C}^{N_{S}\times N_{S}}$ contain the inverse of the respective polarizabilities \cite{pulido2017polarizability}, whereas the diagonal elements of the $\mathbf{G}$ matrices in the off-diagonal blocks are zeros. It is worth highlighting that the negative signs in the matrices modeling the mutual interactions between the metamaterial elements and the coupling slots (i.e., $\mathbf{G}^{mm,MS}_{xx}$, $\mathbf{G}^{em,MS}_{zx}\in\mathbb{C}^{N_{M}\times N_{S}}$ and $\mathbf{G}^{mm,SM}_{yx}$, $\mathbf{G}^{me,SM}_{yz}\in\mathbb{C}^{N_{S}\times N_{M}}$) account for the opposite orientations of the magnetic dipole moments of the coupling slots when observed from below and above the slot aperture, as depicted in Fig. \ref{Fig1_Schematic}(c) (see the purple arrows). More specifically, the sign difference is a result of the boundary conditions imposed at the slot apertures \cite{collin1960field,jackson1999classical}.

Once (\ref{DDA_eq_singletile}) is solved for the dipole moments via matrix inversion, the radiation pattern can be computed as the superposition of the contributions from the individual dipoles representing the metamaterial elements, i.e., $\mathbf{m}_{x}^{M}$, $\mathbf{m}_{y}^{M}$ and $\mathbf{p}_{z}^{M}$. The far-fields are computed by multiplying the dipole moments by the free-space Green’s function and a factor of $-2$ arising from image theory \cite{collin1960field}.

Using the computed dipole moments for the coupling slots (i.e., $\mathbf{m}_{x}^{S}$), the scattering parameters at the input port of the feed waveguide (i.e., $x=\pm l_{feed}/2$) can also be calculated. Specifically, the scattering parameters at ports $1$ and $2$ are obtained from the ratio of the amplitudes of the forward- and backward-propagating waves. For instance, $S_{11}^{tile}$ and $S_{21}^{tile}$ can be computed as 
\begin{equation} \label{sparam}
\begin{aligned}
S_{11}^{tile} &= \frac{A^{-}\left(x=-l_{feed}/2\right)}{A^{+}_{0}e^{j\beta_{10}l_{feed}/2}}, \\
S_{21}^{tile} &= \frac{A^{+}_{0}e^{-j\beta_{10}l_{feed}/2}+A^{+}\left(x=l_{feed}/2\right)}{A^{+}_{0}e^{j\beta_{10}l_{feed}/2}},
\end{aligned}
\end{equation}
where $A^{+}$ and $A^{-}$ represent the amplitudes of the forward- and backward-propagating waves in the feed waveguide when port 1 is excited, respectively \cite{pozar2009microwave}. $\beta_{10}$ represents the propagation constant of TE$_{10}$ mode in the feed waveguide, and $A^{+}_{0}$ is the amplitude of the incident wave when port 1 is excited. The amplitudes, i.e., $A^{+}$ and $A^{-}$ in (\ref{sparam}), can be calculated as the sum of the field scattered by each element \cite{jackson1999classical}.


\subsection{Model for a Multi-tile Metasurface Excited by a Power Dividing Network} \label{multi_port_analysis}

Having developed the model of the single-tile metasurface in (\ref{DDA_eq_singletile}), we propose a model for multi-tile metasurfaces consisting of an array of the single-tile waveguide-fed metasurfaces---without a power dividing feeder---shown in Fig. \ref{Fig1_Schematic}(a). Assuming an array of $U$ metasurface tiles, the coupled dipole model for the multi-tile system can be written as
\begin{equation} \label{DDA_eq_multitile}
\begin{aligned}
\begin{bmatrix*}[l]
\mathbf{G}^{}_{ms,1} & \mathbf{G}^{}_{fs,12} & \dots & \mathbf{G}^{}_{fs,1U} \\
\mathbf{G}^{}_{fs,21} & \mathbf{G}^{}_{ms,2} & \dots & \mathbf{G}^{}_{fs,2U} \\
\vdots & \vdots & \ddots & \vdots \\
\mathbf{G}^{}_{fs,U1} & \mathbf{G}^{}_{fs,U2} & \dots & \mathbf{G}^{}_{ms,U}
\end{bmatrix*}
\begin{bmatrix*}[l]
\mathbf{J}_{ms,1} \\
\mathbf{J}_{ms,2} \\
\vdots \\
\mathbf{J}_{ms,U}
\end{bmatrix*}=
\begin{bmatrix*}[l]
\mathbf{F}^{}_{ms,1} \\
\mathbf{F}^{}_{ms,2} \\
\vdots \\
\mathbf{F}^{}_{ms,U}
\end{bmatrix*},
\end{aligned}
\end{equation}
where $\mathbf{G}_{ms,u}$ represents the block matrix in (\ref{DDA_eq_singletile}) corresponding to the $u$-th metasurface. 

The off-diagonal elements of $\mathbf{G}_{fs,uv}$ correspond to the free-space Green’s functions that model the mutual interactions between metamaterial radiators in the $u$-th and $v$-th metasurfaces via their radiated fields, while the diagonal entries are set to zero. In this manner, dipolar interactions between metamaterial radiators across different metasurface tiles are incorporated, enabling accurate analysis of the multi-tile system. 

\begin{figure}[!b]
\centering
\includegraphics[width=3.25in]{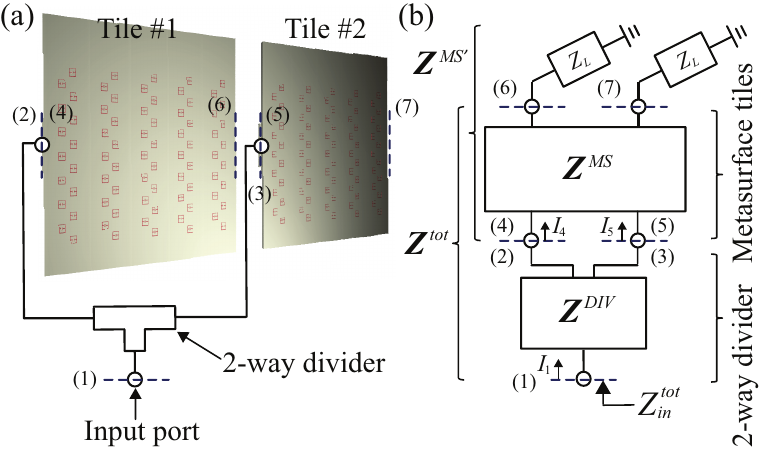}
\caption{(a) Example schematic of a multi-tile system with two metasurface tiles fed by a two-way power divider. Port 1 of the power divider is used to excite the system. (b) Network representation of the multi-tile metasurface shown in Fig. \ref{Fig2_Schematic_Segmentation}(a), with port indices labeled. Dashed lines indicate the reference planes at which the Z- or S-parameters are measured. Ports $6$ and $7$ are terminated with matched loads (i.e., $Z_{L}=Z_{0}$).}
\label{Fig2_Schematic_Segmentation}
\end{figure}


By solving the coupled matrix equation in (\ref{DDA_eq_multitile}), the dipole moments within each metasurface (i.e., $\mathbf{J}_{ms,u}$) can be obtained for the given excitation fields (i.e., $\mathbf{F}_{inc,u}$). The overall radiation pattern of the multi-tile metasurface system can then be computed using free-space Green’s functions, following the approach introduced in subsection \ref{cdm_model}. It should be noted at this point that the size of the matrix equation in (\ref{DDA_eq_multitile}) scales linearly with the number of metamaterial radiating elements. Consequently, the computational burden can remain relatively moderate---even for electrically large aperture systems---when compared to full-wave numerical techniques, as will be discussed in later sections.



The computation of the S-parameters of a multi-tile metasurface system is essential to analyze its performance. The S-parameters of the multi-tile metasurfaces---without a feed network---can be obtained using the dipole moments in (\ref{DDA_eq_multitile}) and calculating the ratio of the forward- and backward-propagating waves, similar to (\ref{sparam}). In this manner, the metasurface structure can be characterized as a $2U$-port microwave network with corresponding scattering matrix $\mathbf{S}^{MS}\in\mathbb{C}^{2U \times 2U}$.

Next, we connect the metasurface tiles to a power-dividing network and determine the overall S-parameters of the multi-tile metasurface system measured at the input port of the power divider. To this end, we propose a method that combines the dipole model in (\ref{DDA_eq_multitile}) with a multiport network analysis framework. Among the existing approaches for multiport-network cascading \cite{chadha1981segmentation,sorrentino2003planar,frei2008multiport}, the segmentation technique reported in \cite{chadha1981segmentation} is employed because it allows the cascading operation to be represented in terms of propagating modes, which is sufficient for the analysis in this work. The detailed procedure for computing the overall network response is presented in the following sections. Specifically, the individual metasurface tile is modeled as a $2U$-port microwave network with impedance matrix $\mathbf{Z}^{MS}\in\mathbb{C}^{2U \times 2U}$, obtained through conversion from the scattering matrix $\mathbf{S}^{MS}$. The metasurface network is then connected to a power divider network characterized by its impedance matrix $\mathbf{Z}^{DIV}\in\mathbb{C}^{(U+1)\times (U+1)}$, which provides excitation to the overall system. Using these impedance matrices, the metasurface network and the power divider are cascaded via the segmentation technique to obtain the impedance matrix of the complete system, denoted by $\mathbf{Z}^{tot}$. In this manner, the proposed model in (\ref{DDA_eq_multitile}) captures the mutual coupling between neighboring metasurface tiles, thereby enabling the prediction of \textit{all} entries of $\mathbf{S}^{MS}$. It is worth noting that such a prediction is possible only by accounting for the mutual interactions among constituent elements, including interactions between metamaterial radiators through guided modes and near-field in free space, as well as interactions between metamaterial radiators and coupling slots through guided modes.



Note that the proposed method---combining the dipole model in (\ref{DDA_eq_multitile}) with the multi-port network analysis technique---for computing the overall S-parameter is fully general and is not limited to multi-tile systems with a specific number of tiles or ports. However, to present the complete formalism and to clearly illustrate the proposed approach, we consider here a multi-tile metasurface system excited by a two-way power divider, as shown in Fig. \ref{Fig2_Schematic_Segmentation}(a). The impedance matrix of the overall system, shown in Fig. \ref{Fig2_Schematic_Segmentation}(a), can be obtained using \cite{chadha1981segmentation}
\begin{equation} \label{Ztot}
\begin{aligned}
\mathbf{Z}^{tot} &= \mathbf{Z}_{pp}^{} + \left(\mathbf{Z}_{pq}^{}-\mathbf{Z}_{pr}^{}\right)\left(\mathbf{Z}_{qq}+\mathbf{Z}_{rr}^{}
\right)^{-1} 
\left(\mathbf{Z}_{rp}-\mathbf{Z}_{qp}^{}\right)
\end{aligned}
\end{equation}
where the entries of the impedance matrices are given by
\begin{equation} \label{Zmatrices_1}
\begin{aligned}
\mathbf{Z}_{pp}&=
\begin{bmatrix}
Z_{11}^{DIV} & 0 & 0 \\
0            & Z_{66}^{MS} & Z_{67}^{MS} \\
0            & Z_{76}^{MS} & Z_{77}^{MS} \\
\end{bmatrix},
\end{aligned}
\end{equation}
and
\begin{equation} \label{Zmatrices_2}
\begin{aligned}
\mathbf{Z}_{pq}&=
\begin{bmatrix}
0           & 0 \\
Z_{64}^{MS} & Z_{65}^{MS} \\
Z_{74}^{MS} & Z_{75}^{MS} \\
\end{bmatrix}, \quad 
\mathbf{Z}_{pr}=
\begin{bmatrix}
Z_{12}^{DIV} & Z_{13}^{DIV} \\
0           & 0 \\
0           & 0 \\
\end{bmatrix}, \\
\mathbf{Z}_{qp}&=
\begin{bmatrix}
0 & Z_{46}^{MS} & Z_{47}^{MS} \\
0 & Z_{56}^{MS} & Z_{57}^{MS} \\
\end{bmatrix}, \quad
\mathbf{Z}_{qq}=
\begin{bmatrix}
Z_{44}^{MS} & Z_{45}^{MS} \\
Z_{54}^{MS} & Z_{55}^{MS} \\
\end{bmatrix}, \quad \\
\mathbf{Z}_{rp}&=
\begin{bmatrix}
Z_{21}^{DIV} & 0 & 0 \\
Z_{31}^{DIV} & 0 & 0 \\
\end{bmatrix}, \quad
\mathbf{Z}_{rr}=
\begin{bmatrix}
Z_{22}^{DIV} & Z_{23}^{DIV} \\
Z_{32}^{DIV} & Z_{33}^{DIV} \\
\end{bmatrix}.
\end{aligned}
\end{equation}
Note that the subscript $p \in \{1,6,7\}$ represents the port indices of the overall metasurface system, while the subscripts $q \in \{4,5\}$ and $r \in \{2,3\}$ represent the indices of the internal ports used to connect the divider network and metasurface tiles.

Having calculated the impedance parameter of the overall structure using (\ref{Ztot}) along with (\ref{Zmatrices_1})--(\ref{Zmatrices_2}), the input impedance seen at the input port of the power divider is calculated as
\begin{equation} \label{Zin}
\begin{aligned}
Z_{in}^{tot} = \frac{\text{det}\left(\mathbf{Z}^{tot}+\mathbf{Z}_{L}\right)}{\text{det}\left(\mathbf{Z}_{d}+\mathbf{Z}_{L}\right)},
\end{aligned}
\end{equation}
where $\mathbf{Z}_{L}=\mathrm{diag}\left(0,Z_{L},Z_{L},Z_{L}\right)$, and $Z_{L}$ represents the impedance of the terminating load. Throughout this work, we assume that the waveguides are terminated with matched loads; therefore, $Z_{L}=Z_{0}$ where $Z_{0}$ represents the wave impedance of TE$_{10}$ mode. $\mathbf{Z}_{d}$ is given by
\begin{equation} \label{Z_d}
\begin{aligned}
\mathbf{Z}_{d} &=
\begin{bmatrix}
1 & [\mathbf{Z}^{tot}]_{1,6} & [\mathbf{Z}^{tot}]_{1,7} \\
0 & [\mathbf{Z}^{tot}]_{6,6} & [\mathbf{Z}^{tot}]_{6,7} \\
0 & [\mathbf{Z}^{tot}]_{7,6} & [\mathbf{Z}^{tot}]_{7,7} \\ 
\end{bmatrix},
\end{aligned}
\end{equation}
where $[\cdot]_{i,j}$ denotes the entry of a matrix in the $i$th row and $j$th column. In this manner, the coupling between the power divider and metasurface tiles can be considered for the computation of the input impedance of the overall multi-tile system. Once the input impedance $Z_{in}^{tot}$ is obtained, the S-parameters can be obtained using the network parameter conversion \cite{pozar2009microwave}.

From (\ref{Ztot}) and (\ref{Zin}), it can be observed that the proposed method for computing the scattering parameters of a multi-tile metasurface system is general and does not depend on the specific design parameters of the power divider or individual metasurface tiles, such as the number of tiles or the geometry of the constituent elements. Although the example in this subsection considers two metasurface tiles fed by a two-way power divider, the proposed method can be readily extended to systems with an arbitrary number of tiles excited by \textit{any} practical power-dividing network. Since the entries of the impedance matrices $\mathbf{Z}^{MS}$ and $\mathbf{Z}^{DIV}$ are required to define the matrices in (\ref{Zmatrices_1}) and (\ref{Zmatrices_2}), and matrix operations are subsequently performed to obtain (\ref{Ztot}) and (\ref{Zin}), the validity of the proposed approach depends on accurate determination of the impedance parameters characterizing both the metasurface tiles and the power divider. It should be noted that, as will be demonstrated in subsequent subsections, $\mathbf{Z}^{DIV}$ is determined using full-wave simulations.

Next, the amplitudes of the voltage and current waves injected into the metasurface tiles (i.e., $I_{4}^{+}$, $I_{5}^{+}$ and $V_{4}^{+}$, $V_{5}^{+}$ in Fig. \ref{Fig2_Schematic_Segmentation}(b)) should be determined for a given input current amplitude $I_{1}$. These voltage and current waves are required for accurate prediction of the radiated power of the multi-tile metasurface system, as will be discussed in this subsection. To compute the voltage and current waves, we begin by calculating the impedance matrix of the metasurface tiles terminated with matched loads (i.e., $Z_{L} = Z_{0}$ in Fig. \ref{Fig2_Schematic_Segmentation}(a)). The corresponding microwave network representation is shown in Fig. \ref{Fig2_Schematic_Segmentation}(b). To derive the relationship between the voltages and currents at ports $4$ and $5$ of the resulting network (i.e., the metasurface tiles with ports $6$ and $7$ terminated with matched loads), we introduce the following matrices for notational simplicity. Specifically, we partition the impedance matrix $\mathbf{Z}^{MS}$ into $2$-by-$2$ block matrices as
\begin{equation} \label{partitioned_matrix1}
\begin{aligned}
\begin{bmatrix}
\mathbf{V}_{A} \\
\mathbf{V}_{B}
\end{bmatrix}=
\begin{bmatrix}
\mathbf{Z}_{AA}^{MS} & \mathbf{Z}_{AB}^{MS} \\
\mathbf{Z}_{BA}^{MS} & \mathbf{Z}_{BB}^{MS} \\
\end{bmatrix}
\begin{bmatrix}
\mathbf{I}_{A} \\
\mathbf{I}_{B} \\
\end{bmatrix}
\end{aligned}
\end{equation}
where
\begin{equation} \label{partitioned_impedance_matrices1}
\begin{aligned}
\mathbf{Z}_{AA}^{MS}&=
\begin{bmatrix}
Z_{44}^{MS} & Z_{45}^{MS} \\
Z_{54}^{MS} & Z_{55}^{MS} \\
\end{bmatrix}, \quad 
\mathbf{Z}_{AB}^{MS}=
\begin{bmatrix}
Z_{46}^{MS} & Z_{47}^{MS} \\
Z_{56}^{MS} & Z_{57}^{MS} \\
\end{bmatrix}, \\
\mathbf{Z}_{BA}^{MS}&=
\begin{bmatrix}
Z_{64}^{MS} & Z_{65}^{MS} \\
Z_{74}^{MS} & Z_{75}^{MS} \\
\end{bmatrix}, \quad 
\mathbf{Z}_{BB}^{MS}=
\begin{bmatrix}
Z_{66}^{MS} & Z_{67}^{MS} \\
Z_{76}^{MS} & Z_{77}^{MS} \\
\end{bmatrix},
\end{aligned}
\end{equation}
and
\begin{equation} \label{partitioned_voltages_currents}
\begin{aligned}
\mathbf{V}_{A}&=
\begin{bmatrix}
V_{4} & V_{5} \\
\end{bmatrix}^{T}, \quad 
\mathbf{V}_{B}=
\begin{bmatrix}
V_{6} & V_{7} \\
\end{bmatrix}^{T}, \\
\mathbf{I}_{A}&=
\begin{bmatrix}
I_{4} & I_{5} \\
\end{bmatrix}^{T}, \quad 
\mathbf{I}_{B}=
\begin{bmatrix}
I_{6} & I_{7} \\
\end{bmatrix}^{T}.
\end{aligned}
\end{equation}

By applying the boundary condition (i.e., $\mathbf{V}_{B}=-Z_{0}\mathbf{I}_{B}$) and performing algebraic computations, we obtain the following matrix representation of an equivalent two-port network modeling the terminated metasurface tiles $\mathbf{Z}^{MS'}\in\mathbb{C}^{2 \times 2}$, with input ports $4$ and $5$, given by
\begin{equation} \label{partitioned_matrix}
\begin{aligned}
\mathbf{Z}^{MS'}=\mathbf{Z}_{AA}^{MS}-\mathbf{Z}_{AB}^{MS}\left(\mathbf{Z}_{BB}^{MS}+Z_{0}\mathbf{U}\right)^{-1}\mathbf{Z}_{BA}^{MS}
\end{aligned}
\end{equation}
where $\mathbf{U}$ represents the identity matrix.

Next, we find the relation between the current injected into the input port of the power divider (i.e., port $1$ in Fig. \ref{Fig2_Schematic_Segmentation}(b)) and the current amplitudes flowing into the input ports of the metasurface tiles (i.e., ports $4$ and $5$ in Fig. \ref{Fig2_Schematic_Segmentation}(b)). Specifically, the amplitudes of the currents can be calculated using
\begin{equation} \label{Iq}
\begin{aligned}
\mathbf{I}_{q}^{} &= \mathbf{Z}_{qp'}^{'}\mathbf{I}_{p'}^{'},
\end{aligned}
\end{equation}
where $\mathbf{Z}_{qp'}^{'}=\left(\mathbf{Z}_{qq}^{MS'}+\mathbf{Z}_{rr}^{}\right)^{-1}\mathbf{Z}_{rp'}^{MS'}$ is the matrix relating the amplitudes of the total currents measured at the internal ($q\in \{4,5\}$) and external ports ($p'\in \{1\}$) of the terminated metasurface tiles. The elements of the matrices $\mathbf{Z}_{qq}^{MS'}$ and $\mathbf{Z}_{rp}^{MS'}$ are calculated as
\begin{equation} \label{partitioned_impedance_matrices}
\begin{aligned}
\mathbf{Z}_{qq}^{MS'}&=
\begin{bmatrix}
Z_{44}^{MS'} & Z_{45}^{MS'} \\
Z_{54}^{MS'} & Z_{55}^{MS'} \\
\end{bmatrix}, \quad
\mathbf{Z}_{rp}^{MS'}=
\begin{bmatrix}
Z_{21}^{DIV} \\
Z_{31}^{DIV} \\
\end{bmatrix}.
\end{aligned}
\end{equation}
The amplitudes of the total voltage waves are given by
\begin{equation} \label{Vq}
\begin{aligned}
\mathbf{V}_{q}^{} &= \mathbf{Z}_{qq}^{}\mathbf{Z}_{qp}^{'}\mathbf{I}_{p}^{'}.
\end{aligned}
\end{equation}
Using (\ref{Iq}) and (\ref{Vq}), the amplitude of the voltage wave injected into the input port of the metasurface tiles can be obtained, given by
\begin{equation} \label{Vinc}
\begin{aligned}
\mathbf{V}_{q}^{+} &= \begin{bmatrix}
V_{4}^{+} \\ V_{5}^{+} \\
\end{bmatrix}= \frac{1}{2}\left(\mathbf{V}_{q}^{}+Z_{}\mathbf{I}_{q}^{}\right),
\end{aligned}
\end{equation}
The computed voltage amplitudes in (\ref{Vinc}) can be used to determine the magnitude and phase of the fields exciting the metasurface tiles (i.e., $\mathbf{F}_{ms,1}$ and $\mathbf{F}_{ms,2}$ in (\ref{DDA_eq_multitile})). Note that (\ref{Vinc}) can also be derived from the S-parameters of the networks comprising the multi-tile metasurface system; the detailed derivation is provided in Appendix A.

For a given power incident on the input port of the divider, the amplitude of the input current at port $1$, denoted as $I_{1}$, (i.e., $\mathbf{I}_{p'}^{'}=I_{1}$), is given by
\begin{equation} \label{I1}
I_{1} = \sqrt{\frac{2P^{i}}{\text{Re}\{1/Z_{0}\}}},
\end{equation}
where $P^{i}$ is the power incident on port $1$ of the divider.

Once the voltage wave entering the input port of the overall metasurface system is obtained from (\ref{Vq}) and (\ref{I1}), the amplitudes of the incident magnetic fields in (\ref{DDA_eq_singletile}) (i.e., $\mathbf{H}^{inc}_{x}$) need to be determined. It should be noted that the relationships among voltages, currents, and electromagnetic fields in waveguides are not uniquely defined \cite{pozar2009microwave}. Therefore, by assuming that the dominant TE$_{10}$ mode is the only propagating mode in the waveguide, we define equivalent voltages and currents for TE$_{10}$ mode using appropriate proportionality constants. The constants are selected such that the ratio of voltage to current equals the wave impedance, while the product of voltage and current corresponds to the power flow along the waveguide \cite{pozar2009microwave,harrington1961timeharmonic}. Accordingly, we employ the following relations, expressed as
\begin{equation} \label{Equivalent_V_I}
\begin{aligned}
V^{\pm}_{0} = A^{\pm}_{0},\quad I^{\pm}_{0} = \frac{1}{Z^{}_{0}}A^{\pm}_{0},
\end{aligned}
\end{equation}
where $A^{+}_{0}$ and $A^{-}_{0}$ represent the amplitudes of the forward- and backward-propagating waves within the waveguide, respectively. Also, $V^{\pm}_{0}$ and $I^{\pm}_{0}$ are the corresponding voltage and current amplitudes, respectively. Specifically, the entries of the incident magnetic field vector are given by \cite{jackson1999classical}
\begin{equation} \label{Hx_inc}
\begin{aligned}
H^{+}_{x0}=A^{+}_{0}\frac{j\pi\sqrt{2}}{a\sqrt{ab}Z_{0}\beta_{10}}\cos\left(\frac{\pi y_{n_m}}{a}\right)e^{-j\beta^{}_{10}x_{n_m}},
\end{aligned}
\end{equation}
where $(x_{n_m},y_{n_m})$ denotes the location of the ${n_m}$-th metamaterial element, as illustrated in Fig. \ref{Fig1_Schematic}(b). Here, $a$ and $b$ represent the width and height of the rectangular waveguide, respectively.

\section{Full-wave Verification} \label{full_wave_verification}

We begin this subsection by validating the single-tile metasurface model presented in (\ref{DDA_eq_singletile}) through full-wave simulation of a design example in Subsection \ref{single_tile_example}. The verification is then extended to the model of a multi-tile metasurface system fed by a practical two-way power divider in Subsection \ref{multitile_example}. It should be emphasized that the primary goal of this subsection is to validate the proposed models and demonstrate their effectiveness as analysis and design tools for multi-tile metasurface systems, rather than to present metasurface designs fully optimized for specific applications. To this end, the examples considered have overall aperture dimensions on the order of a few wavelengths, selected to balance the total number of mesh elements, convergence, and the computational time and accuracy of the full-wave simulations.


\subsection{Single-tile Metasurface Example} \label{single_tile_example}

\begin{figure}[!t]
\centering
\includegraphics[width=3.5in]{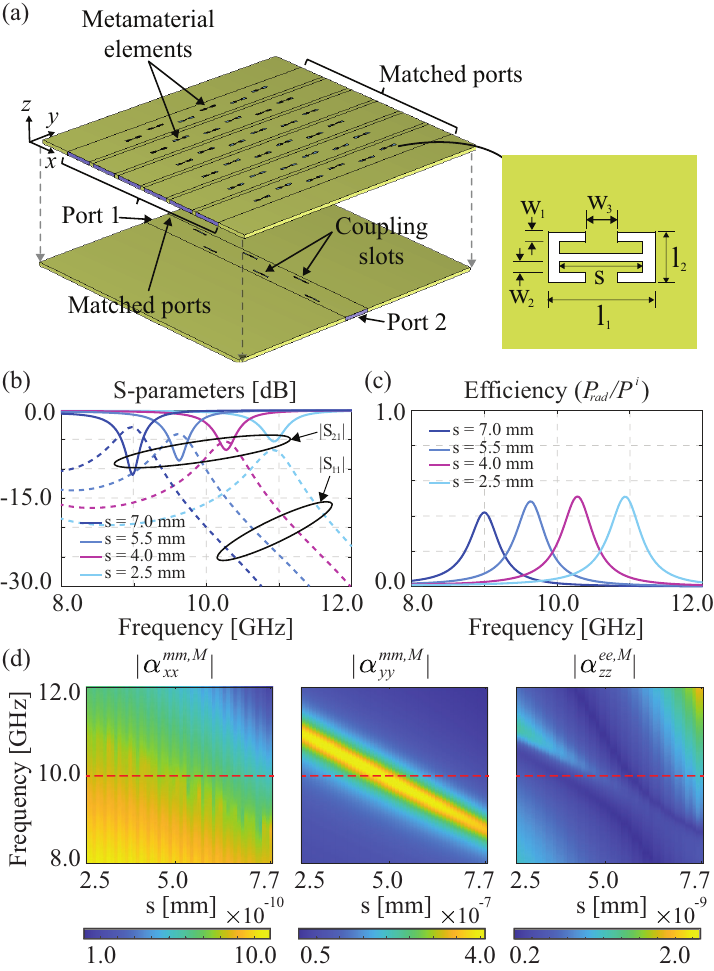}
\caption{(a) Exploded view of the designed metasurface antenna. The radiative metasurface layer consists of 6 rectangular waveguides, each with 6 metamaterial elements. The metasurfaces are fed by a slotted waveguide at the bottom. In the schematic shown in Fig. \ref{Fig3_ResonantPolarizability}(a), the slotted waveguide is embedded in a conducting box for illustration. Yellow indicates the perfect electric conductor. The design parameters are: $w_{1}=0.225$ mm, $w_{2}=0.2$ mm, $w_{3}=2.5$ mm, $l_{1}=8.0$ mm, $l_{2}=1.0$ mm. (b) S-parameters of the metamaterial elements for the selected arm length $s$, and (c) radiated power $P_{rad}$ divided by the excitation power $P^{i}$. (d) The amplitude of the extracted magnetic polarizability ($\alpha_{xx,M}^{mm}$, $\alpha_{yy,M}^{mm}$) and electric polarizability ($\alpha_{zz,M}^{ee}$) of the element depicted in Fig. \ref{Fig3_ResonantPolarizability}(a). The polarizabilities are shown as functions of frequency and gap size $s$. Red, dashed curves are used to indicate the operating frequency of $f_{op}=10.0$ GHz.}
\label{Fig3_ResonantPolarizability}
\end{figure}

We consider the schematic of a single-tile metasurface antenna shown in Fig. \ref{Fig3_ResonantPolarizability}(a). As depicted in \ref{Fig3_ResonantPolarizability}(a), the single-tile antenna consists of $N_{S}=6$ rectangular waveguide-fed metasurfaces, each excited by fields coupled through a slot located between the bottom wall of the metasurface and the top wall of a feeding rectangular waveguide. The coupling slots are excited by the guided modes propagating along the feed waveguide. In this design example, identical rectangular waveguides filled with Rogers 3003 substrate ($\epsilon_r=3.0$ and $\tan\delta=0.001$) are employed to form both the metasurfaces and the feeding waveguide. The width and height of each waveguide are $a=12.0$ mm and $b=1.52$ mm, respectively. The center-to-center spacing between adjacent waveguides in the radiating metasurface layer is $s_{wg}=14.0$ mm.

Note that, for structural simplicity, the waveguide dimensions in this design example are assumed to be identical. However, the proposed model in (\ref{DDA_eq_singletile}) is general and can be applied to metasurfaces with waveguides with different dimensions. In such cases, weighting factors other than negative unity should be incorporated into the matrix on the left-hand side of (\ref{DDA_eq_singletile}). The flexibility in the waveguide dimensions can improve the applicability of the proposed model, enabling the design of metasurfaces optimized for specific applications.

Given the fixed waveguide dimensions, we design metamaterial radiators etched into the top plate of the waveguides. The geometry and key dimensions of the element are illustrated in Fig. \ref{Fig3_ResonantPolarizability}(a). The element dimensions are selected such that resonance occurs near the operating frequency of $10.0$ GHz for a range of swept gap sizes $s$. The element is offset by $1.25$ mm from the center of the waveguide to achieve a moderate coupling strength between the element and the guided modes.

Figure \ref{Fig3_ResonantPolarizability}(b) shows the magnitude of the S-parameters for a selected set of gap sizes $s$. As shown in Fig. \ref{Fig3_ResonantPolarizability}(b), $|S_{11}|$ and $|S_{21}|$ remain above $-10$ dB and $-5$ dB at resonance, respectively, thus indicating that the element is not strongly coupled to the waveguide modes. Figure \ref{Fig3_ResonantPolarizability}(c) shows the radiation efficiency of the element as a function of the gap size, computed as the ratio of radiated power $P_{rad}$ to the excitation power $P^{i}=0.5$ W. As shown in Fig. \ref{Fig3_ResonantPolarizability}(c), approximately $40\%$ of the injected power is radiated into free space. Note that, for practical designs, the geometry of the element may need to be modified to adjust the efficiency shown in Fig. \ref{Fig3_ResonantPolarizability}(c), and the number of elements may need to be carefully chosen to achieve high aperture efficiency of a single-tile metasurface \cite{smith2017analysis}. Here, we focus on validating the proposed model for the single-tile metasurface rather than optimizing specific antenna parameters, including aperture efficiency.

It should be noted that the performance of a metasurface depends heavily on the electromagnetic properties---particularly the radiation characteristics---of its constituent metamaterial elements. Consequently, tailoring the geometry of the elements to achieve a desired electromagnetic response is generally regarded as a critical step in metasurface design \cite{smith2017analysis}. In this work, we focus on validating the proposed model in (\ref{DDA_eq_singletile}), rather than presenting a metasurface design in which the key performance metrics are fully optimized through detailed tuning of the radiation properties of the metamaterial elements.

For the characterization of the element, we calculate the effective polarizabilities (i.e., $\alpha_{xx}^{mm,M}$, $\alpha_{yy}^{mm,M}$, $\alpha_{zz}^{ee,M}$) using the extraction technique outlined in \cite{pulido2017polarizability}. Specifically, we computed the dipole moments using $\bar{p} = \frac{\epsilon_{0}}{2}\iint \left(x\hat{x}+y\hat{y}\right)\times\bar{J}_{m} ds$ and $\bar{m} = \frac{1}{j\omega\mu_0}\iint \bar{J}_{m} ds$ with $\bar{J}_{m}=-\hat{z}\times\bar{E}^{ap}$ representing the surface magnetic current \cite{collin1960field}. The aperture electric field, i.e., $\bar{E}^{ap}$, was calculated using a full-wave solver, CST Microwave Studio. Here, $\epsilon_{0}$ and $\mu_{0}$ represent the free-space electric permittivity and permeability, respectively. $\omega$ represents the angular frequency. The obtained dipole moments are then divided by the incident fields at the element center to compute the polarizabilities.

To characterize the element, the retrieval technique was applied over a range of gap sizes $s$ from $2.5$ mm to $7.7$ mm. The extracted polarizabilities as functions of frequency and gap size are plotted in Fig. \ref{Fig3_ResonantPolarizability}(d). As shown in Fig. \ref{Fig3_ResonantPolarizability}(d), the magnetic polarizability along the $y$-axis (i.e., $\alpha_{yy}^{mm,M}$) exhibits a dominant Lorentzian resonance, while the other components remain small and relatively unchanged across the swept frequency and gap range. Note that, in this design example, we use the polarizabilities at the operating frequency of $f_{op}=10.0$ GHz for the design. Nevertheless, the extracted polarizabilities are plotted over a broader frequency range in Fig. \ref{Fig3_ResonantPolarizability}(d) to illustrate the resonant behavior of the element and to validate the proposed model through S-parameter calculations, which are presented later in this subsection.


As a next step, we design the coupling slots and the feed waveguide. We consider the rectilinear slots with a fixed width of $0.5$ mm, positioned with an offset of $4.0$ mm from the center of the feed waveguide, as shown in Fig. \ref{Fig3_ResonantPolarizability}(a). The slot dimensions are chosen to be electrically small, and the axial ratio is chosen to be sufficiently large to ensure that the coupling slots can be well-modeled as magnetic dipoles oriented along their principal axes. Under these design considerations, each slot is characterized by its effective magnetic polarizability, $\alpha_{xx}^{mm,S}$.


\begin{figure}[!t]
\centering
\includegraphics[width=3.5in]{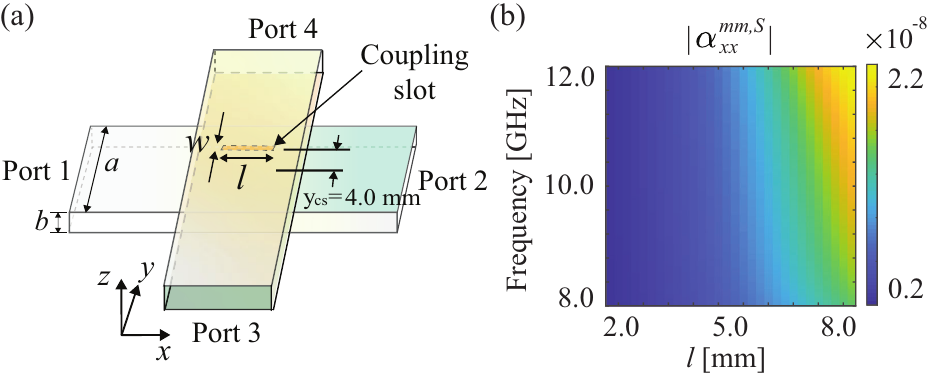}
\caption{(a) Simulation setup for the retrieval of the effective polarizability $\alpha_{xx}^{mm,S}$ of a rectilinear coupling slot. Two rectangular waveguides are coupled via the slot. Port 1 is used to excite the structure, while the rest is terminated with the matched load. The center of the slot is $x_{cs}=0.0$ mm and $y_{cs}=4.0$ mm. The width of the slot is fixed to be $0.5$ mm, while the length is varied from $2.0$ to $8.0$ mm. (b) The magnitude of the extracted polarizability of the coupling slot depicted in Fig. \ref{Fig4_CouplingSlotPolarizability}(a).}
\label{Fig4_CouplingSlotPolarizability}
\end{figure}


For polarizability retrieval of the coupling slot, we consider the schematic of two rectangular waveguides coupled through a rectilinear slot positioned at $(x=0,y=y_{cs})$, as depicted in Fig. \ref{Fig4_CouplingSlotPolarizability}(a). In this setup, port $1$ of the lower waveguide is used to excite the structure, while ports $2$, $3$ and $4$ are assumed to be terminated with matched loads. Assuming that $\alpha_{yy}^{mm,S},\alpha_{zz}^{ee,S}\approx 0$, the amplitude of the forward-going wave in the lower (feed) waveguide can be expressed as \cite{collin1960field,pulido2017polarizability}
\begin{equation} \label{VoltageWave}
\begin{aligned}
A^{+}\left(x\right) &= \frac{j\omega Z_{}}{2} \Big(\mu\alpha_{xx}^{mm}H^{+}_{x0}H^{-}_{x0}\Big),
\end{aligned}
\end{equation}
where $\mu$ represents the permeability of a substrate filling the feed waveguide. $H^{+}_{x0}$ and $H^{-}_{x0}$ represent the forward and backward-going incident magnetic fields at the center of the coupling slot in the waveguide, respectively. Using the eigenmode of the rectangular waveguide, (\ref{VoltageWave}) can be expressed in terms of the S-parameter of the feed waveguide, given by
\begin{equation} \label{polarizability_coupling_slot}
\begin{aligned}
\alpha_{xx}^{mm,S} &= \frac{j a^{3}b\beta}{2\pi^2 \cos^{2}\left(\frac{\pi y_{cs}}{a}\right)} \Big(S_{21}+S_{11}-1\Big),
\end{aligned}
\end{equation}
where $S_{11}$ and $S_{21}$ represent the S-parameters measured at the center of the slot (i.e., $x_{cs}=0.0$ mm). It should be noted that (\ref{polarizability_coupling_slot}) is valid only under the condition that $\alpha_{yy}^{mm,S},\alpha_{zz}^{ee,S}\approx 0$. Consequently, the longest principal axis of the slot should be oriented along the $\hat{x}$-direction, and the axial ratio of the slot needs to be sufficiently high. Figure \ref{Fig4_CouplingSlotPolarizability}(b) shows the extracted polarizability of the coupling slot as functions of frequency and slot length, with the length varied from $2.0$ mm to $8.0$ mm. The width of the slot is fixed at $0.5$ mm. As shown in Fig. \ref{Fig4_CouplingSlotPolarizability}(b), the magnitude of the polarizability increases as the slot length increases \cite{collin1960field}. It should be noted that electromagnetic coupling of two regions or waveguides has been studied extensively in the literature \cite{collin1982small,mautz1983admittance,liang1983generalized}, and we here employ the coupled dipole framework to model the coupling of two rectangular waveguides through multiple coupling slots.


We then insert the coupling slots into the common wall of the metasurfaces and the feed waveguide. The lengths of the coupling slots are determined such that the guided modes propagating along the feed waveguide are coupled to the array of metasurfaces as evenly and efficiently as possible ($\sim -15.1$ dB), while also achieving impedance matching at port 1 and reducing $|S_{21}|$. The design process for the feed waveguide is performed using the reduced dipole model of (\ref{DDA_eq_singletile}), in which the dipole moments associated with the metamaterial elements are excluded (see Appendix B). It is worth noting that, in principle, the slot lengths and the geometrical parameters (e.g., gap sizes) of the individual metamaterial elements can be jointly optimized in the design of a single-tile metasurface. However, in this work, the design procedure is decoupled: the feed waveguide geometry (i.e., coupling slot lengths) is determined first, followed by the metasurface design (i.e., metamaterial element geometry). This approach reduces the number of parameters to be optimized in subsequent design stages.


\begin{figure}[!t]
\centering
\includegraphics[width=3.5in]{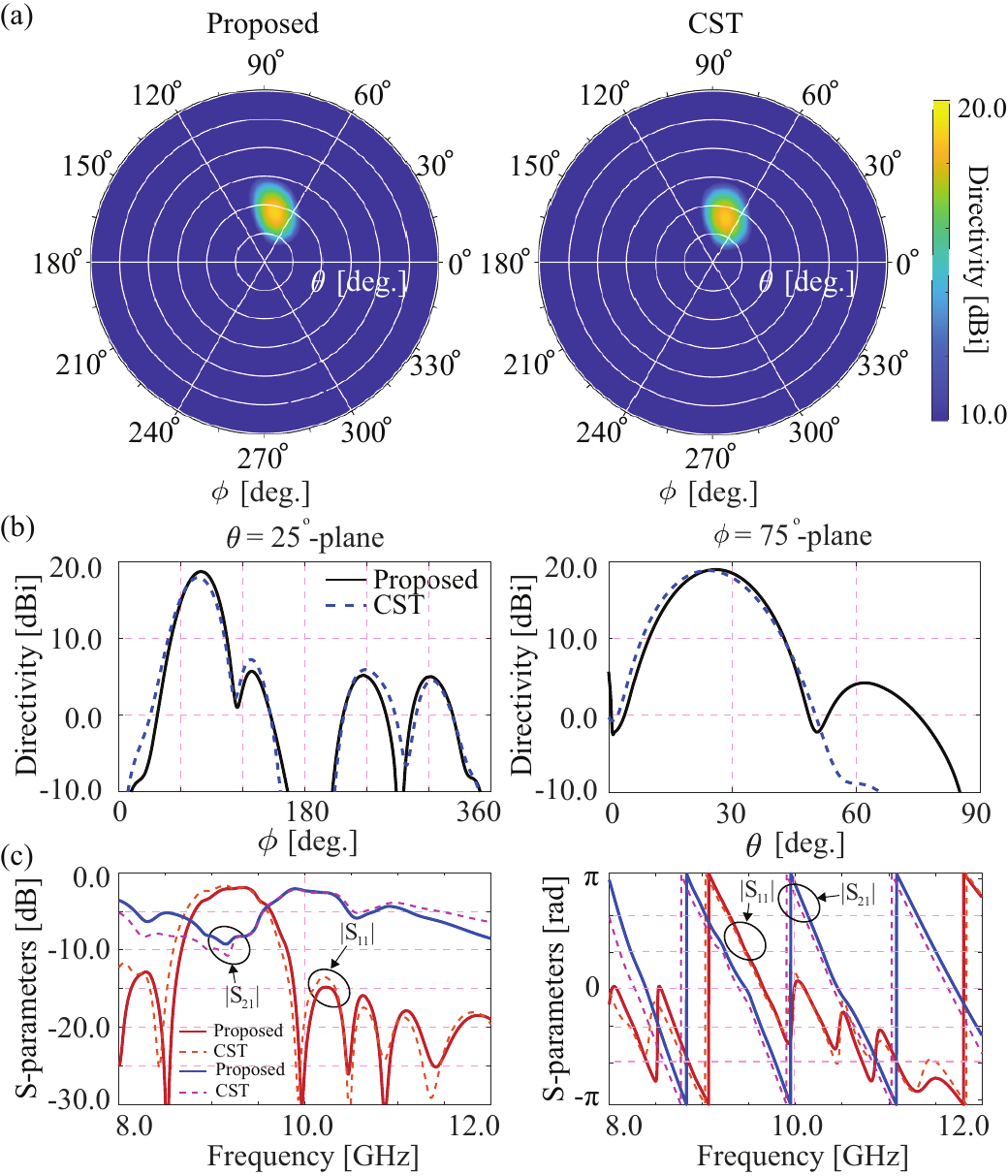}
\caption{(a) Directivity pattern obtained using the proposed model and full-wave analysis. The frequency is $f_{op}=10.0$ GHz. (b) Cross-sectional plots of the directivity patterns at the planes $\theta=25^{\circ}$ (left) and $\phi=75^{\circ}$ (right). (c) The magnitude and phase of S-parameters.}
\label{Fig5_Pattern_6_by_6}
\end{figure}

Next, a desired radiation pattern is synthesized by tuning the geometry of the individual metamaterial elements. To this end, $N_{M}=6$ metamaterial radiators are loaded into each waveguide to form a rectangular waveguide-fed metasurface. The center-to-center distance between adjacent elements within each waveguide is $14.0$ mm, and that between adjacent waveguides is set to $14.0$ mm. Using the relation between the element geometry (i.e., gap size) and its polarizability, as shown in Fig. \ref{Fig3_ResonantPolarizability}(d), a set of gap sizes, denoted by $\mathcal{S}$, is determined to produce the desired radiation pattern at the operating frequency $f_{op}=10.0$ GHz. For this design example, a single-tile metasurface is synthesized to generate a directive beam toward the target direction $\left(\phi_{tar},\theta_{tar}\right)=\left(75^{\circ},25^{\circ}\right)$.

To determine the set of gap sizes, we employ the pattern search technique \cite{audet2002analysis} with a cost function, defined as
\begin{equation} \label{cost_function}
\begin{aligned}
\text{cost} = -\Big[G_{dB,single}\left(f_{op},\mathcal{S},\phi_{tar},\theta_{tar}\right) + SLL_{dB}\left(f_{op},\mathcal{S}\right)\Big],\\
\end{aligned}
\end{equation}
where $G_{dB,single}$ and $SLL_{dB}$ represent, respectively, the gain of the single tile in the target direction and the sidelobe level in decibel (dB) scale, when port $1$ of the metasurface tile is excited. The gain of the single-tile metasurface is computed as the product of the directivity and efficiency, where the efficiency is defined as the radiated power divided by the excitation power, i.e., $P_{rad}/P^{i}$. It should be noted that, for the multi-tile system design, the initial slot lengths were uniformly set to $5.0$ mm to demonstrate the robustness of the proposed design approach.

Figure \ref{Fig5_Pattern_6_by_6}(a) shows predicted and simulated directivity patterns at the operating frequency of $f_{op}=10.0$ GHz. In the predicted pattern, the main beam with its peak amplitude of $19.1$ dB is formed at $\left(\phi^{}_{}=77.8^{\circ}, \theta^{}_{}=26.5^{\circ}\right)$. The main beam of the simulated pattern is located at $\left(\phi^{}_{}=74.2^{\circ}, \theta^{}_{}=23.9^{\circ}\right)$, with its peak amplitude of $18.9$ dB. Though slightly offset from the target direction, the predicted and simulated beams are generated near the intended direction, also demonstrating excellent agreement in the shape of the patterns. In Fig. \ref{Fig5_Pattern_6_by_6}(b), we plot the cross-sections of the radiation patterns in Fig. \ref{Fig5_Pattern_6_by_6}(a) at the plane of $\phi=\phi_{tar}$ and $\theta=\theta_{tar}$, respectively. As shown in Fig. \ref{Fig5_Pattern_6_by_6}(b), the predicted and simulated patterns exhibit great agreement, thus confirming the effectiveness of the model in (\ref{DDA_eq_singletile}). It should be noted that the predicted directivity pattern on $\phi=75^{\circ}$-plane begins to deviate from the simulated pattern when $\theta>51^{\circ}$. This deviation is attributed to a slight shift in the main beam direction, along with the finite size of the simulated metasurface, which contrasts with the assumption of an infinite conducting plane with metamaterial apertures used in the proposed model. 


Figure \ref{Fig5_Pattern_6_by_6}(c) shows the magnitude and phase of the S-parameters for the designed single-tile metasurface antenna over the frequency range from $8.0$ to $12.0$ GHz. The predicted and simulated S-parameters exhibit close agreement in both magnitude and phase, thereby validating the proposed method for computing the S-parameters in (\ref{sparam}). As will be discussed in subsequent sections, the level of accuracy seen in Fig. \ref{Fig5_Pattern_6_by_6}(c) is essential for analyzing the overall response of multi-tile metasurface systems, as presented in Subsection \ref{multi_port_analysis}. Additionally, as shown in Fig. \ref{Fig5_Pattern_6_by_6}(c), the predicted and simulated $|S_{11}|$ exhibit dips of $-24.8$ dB and $-18.7$ dB, respectively, at the operating frequency $f_{op}=10.0$ GHz, indicating that the input port of the metasurface is matched. 


At this point, it is important to note that the efficiency of the metasurface can be improved by increasing the number of metamaterial elements and/or rectangular waveguide–fed metasurfaces, thereby enabling greater coupling of energy into free space. However, increasing the number of elements and metasurfaces also increases the number of mesh elements required to achieve sufficient accuracy in validating the model in (\ref{DDA_eq_singletile}), which in turn leads to higher computational cost and simulation time. In this design example, the number of metamaterial elements and rectangular waveguide–fed metasurfaces was selected to balance computational efficiency and simulation time, while maintaining a sufficient mesh density to ensure accurate verification of the proposed model in (\ref{DDA_eq_singletile}).



Having verified the validity of the proposed model in (\ref{DDA_eq_singletile}), it is worth emphasizing that the model can be applied to analyze similar metasurface tile structures. For example, a single metasurface tile with an array of metasurfaces, each terminated by metal walls to form shorted waveguides or cavities, can be modeled by incorporating the appropriate Green’s functions into (\ref{DDA_eq_singletile}). Furthermore, it should be noted that coupling slots exhibiting magnetic dipole moments along the $\hat{z}$ axis (i.e., $\mathbf{m}_{z}^{S}$), in addition to those along the $\hat{x}$ axis (i.e., $\mathbf{m}_{x}^{S}$), can be considered to generalize the model.



\subsection{Multi-tile Metasurface Example} \label{multitile_example}


\begin{figure}[!t]
\centering
\includegraphics[width=3.5in]{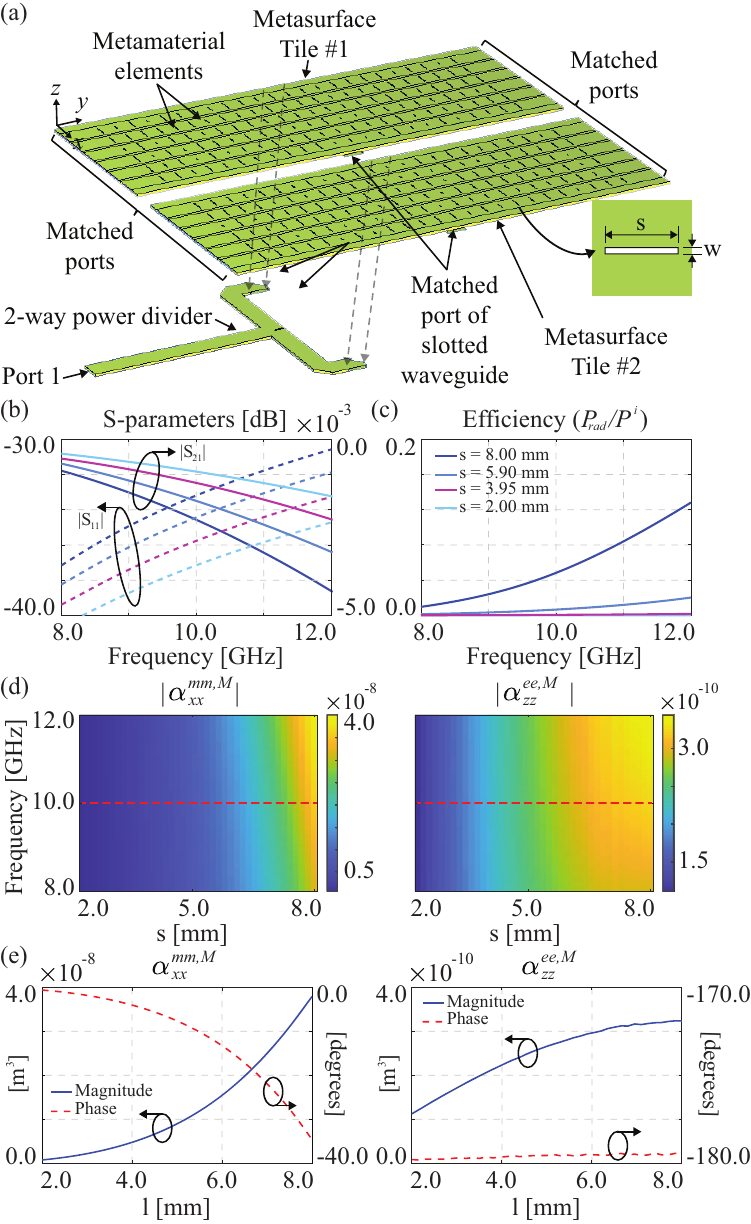}
\caption{(a) Exploded view of the schematic of the designed multi-tile metasurface antenna consisting of two metasurface tiles, fed by a 3-dB power divider. Each metasurface tile consists of $6$ rectangular waveguides, each containing $24$ rectilinear slots. The slotted waveguide in each tile is connected to the power divider to excite the overall multi-tile system. Yellow indicates the perfect electric conductor. The slot width is fixed to be $w=0.5$ mm (inset). (b) S-parameters, and (c) radiated power $P_{rad}$ divided by the excitation power $P^{i}$ of the slot elements for the selected slot length $l$. The excitation power is $P^{i}=0.5$ W. (d) The amplitude of the extracted magnetic polarizability $\alpha_{xx}^{mm,M}$ and electric polarizability $\alpha_{zz}^{ee,M}$ as functions of frequency and slog length $s$. (e) The amplitude and phase of the extracted magnetic polarizability $\alpha_{xx}^{mm,M}$ and electric polarizability $\alpha_{zz}^{ee,M}$ at the target frequency of $f_{op}=10.0$ GHz.}
\label{Fig6_NonresonantPolarizability}
\end{figure}

In this subsection, we validate the proposed model in Subsection \ref{multi_port_analysis} by presenting a design example of a multi-tile metasurface antenna. In the design example, we consider two metasurface tiles excited by RF signals of equal magnitude and phase through a two-way, equal-split power divider. Note that our aim here is to validate the proposed model and demonstrate its effectiveness as an efficient analysis and design tool for multi-tile metasurface systems. To this end, the overall aperture dimensions are carefully selected to limit the total number of mesh elements and mitigate potential convergence issues that may affect the accuracy of the full-wave simulations. For the same reason, rectangular slots are employed as radiators in the metasurface tiles, as each requires significantly fewer mesh elements for accurate modeling than the resonant elements used in Subsection \ref{single_tile_example}. It should be noted that the configuration depicted in Fig. \ref{Fig6_NonresonantPolarizability}(a) is analogous—at the architectural level—to conventional slotted waveguide antennas, which can be analyzed using Elliott’s design procedure \cite{elliott1983improved,elliott1986design}, admittance models \cite{mautz1983admittance,liang1983generalized,harrington1991} and extended circuit models that account for mutual interactions between slots via additional impedance terms \cite{elliott1988design,rengarajan2008optimization}. However, unlike conventional slotted waveguide antennas, where the typical slot spacing is approximately 0.5$\lambda_{0}$ (with $\lambda_{0}$ denoting the free-space wavelength), we consider an array of rectangular slots with a subwavelength spacing. Furthermore, in contrast to approaches that model the coupling between neighboring slots through a limited set of modes (e.g., TE$_{10}$ and TE$_{20}$) \cite{elliott1988design,rengarajan2008optimization}, the proposed framework allows for the incorporation of an \textit{arbitrary} number of higher-order waveguide modes via Green’s functions; in practice, only the first few modes are required. In addition, although rectilinear slots are employed as radiators and coupling irises in this work to reduce computational burden, the proposed framework is sufficiently general to model metasurfaces incorporating arbitrarily shaped coupling slots, resonant metamaterial radiators with complex geometries, or dynamically tunable elements, as envisioned in the literature \cite{yoo2016efficient,sleasman2015dynamic,boyarsky2021electronically}. In such cases, conventional slotted waveguide antenna models are not directly applicable.

Figure \ref{Fig6_NonresonantPolarizability}(a) illustrates the schematic of the designed multi-tile metasurface antenna. Each metasurface tile consists of $6$ rectangular waveguides, each containing $24$ slot radiators inserted into the top wall, resulting in a total of $288$ radiating elements. The slot width is fixed at $0.5$ mm. The effective polarizabilities of the slots were computed using the standard retrieval technique reported in \cite{pulido2017polarizability} using the S-parameters. The S-parameters for selected slot lengths are shown in Fig. \ref{Fig6_NonresonantPolarizability}(b). As observed, $|S_{11}|$ remains small, while $|S_{21}|$ is close to $0.0$ dB, indicating that the slots are weakly coupled to the guided modes over the swept frequencies. Figure \ref{Fig6_NonresonantPolarizability}(c) shows the radiated power spectra for a fixed incident power of $P^{i}=0.5$ W. As shown in Fig. \ref{Fig6_NonresonantPolarizability}(c), the radiated power increases with slot length; however, only a small fraction of the input power is radiated, further confirming the weak coupling behavior of the slots.

Using the computed S-parameters in Fig. \ref{Fig6_NonresonantPolarizability}(b), the effective polarizabilities were extracted as functions of frequency and slot length $l$, as shown in Fig. \ref{Fig6_NonresonantPolarizability}(d). As shown in Fig. \ref{Fig6_NonresonantPolarizability}(d), the magnitude of the extracted magnetic polarizability increases with both frequency and slot length, whereas the electric polarizability remains relatively unchanged. To further illustrate the behavior of the polarizabilities, Fig. \ref{Fig6_NonresonantPolarizability}(e) shows the amplitude and phase of the polarizabilities at the operating frequency as functions of slot length. As shown, the magnitude of the magnetic polarizability $\alpha_{xx}^{mm,M}$ increases with slot length, resulting in an overall phase variation of approximately $40^\circ$ over the range of slot lengths considered. In contrast, the magnitude of the electric polarizability $\alpha_{zz}^{ee,M}$ increases slightly and remains comparatively small, indicating that the contribution of the electric dipole component to the scattered field is small. Furthermore, the phase of the electric polarizability remains nearly constant at $-180.0^{\circ}$, as shown in Fig. \ref{Fig6_NonresonantPolarizability}(e). Note that, since the radiating slot is located at the center of the waveguide, the magnetic dipole moment along the wave propagation direction (i.e., $\alpha_{yy}^{mm,M}$) is not excited and can be neglected in the proposed model in (\ref{DDA_eq_singletile}) and (\ref{DDA_eq_multitile}), as previously reported in \cite{yoo2025toward}.


Note that electrically small, rectilinear slots are employed as metamaterial radiators in the design of the multi-tile metasurface. The weak coupling of the slots to the guided modes, as indicated in Figs. \ref{Fig6_NonresonantPolarizability}(b)-(e), can often be realized using non-resonant metamaterial elements, including the rectilinear slots considered in this example. Such weak coupling enables the use of holographic pattern synthesis techniques \cite{smith2017analysis}, thereby facilitating the design process for waveguide-fed metasurfaces. However, as noted earlier, the phase span provided by each slot is limited to approximately $40^\circ$, which is significantly smaller than that achievable with the resonant metamaterial elements shown in Fig. \ref{Fig3_ResonantPolarizability}(a). This limited phase range may restrict the flexibility of pattern synthesis and should therefore be compensated for, for example, by densely placing the radiating slots at subwavelength spacing. To this end, an inter-element spacing of $p=10.0$ mm is chosen, allowing the phase progression of the guided modes within the waveguides to be exploited for pattern synthesis \cite{smith2017analysis}. It should be noted that such subwavelength spacing may introduce non-negligible mutual coupling between elements, even though the individual slots are non-resonant. Consequently, accounting for mutual coupling is critical in the proposed dipole model in (\ref{DDA_eq_singletile}) and (\ref{DDA_eq_multitile}) for accurate modeling of the metasurfaces.


For the design, the metasurface tiles are assumed to lie on the $xy$-plane with a center-to-center spacing of $110.0$ mm and are connected to a two-way power divider, as shown in Fig. \ref{Fig6_NonresonantPolarizability}(a). Specifically, the output ports of the power divider are connected to the input ports of the metasurface tiles to enable effective excitation of the multi-tile metasurface system. The divider provides an equal split of the injected RF signal with identical phase over the frequency range from $8.0$ GHz to $12.0$ GHz; the detailed design procedure is provided in Appendix C. The remaining ports, including those of the metasurface tiles, are assumed to be terminated with matched loads, as indicated in Fig. \ref{Fig6_NonresonantPolarizability}(a).


\begin{figure}[!t]
\centering
\includegraphics[width=3.5in]{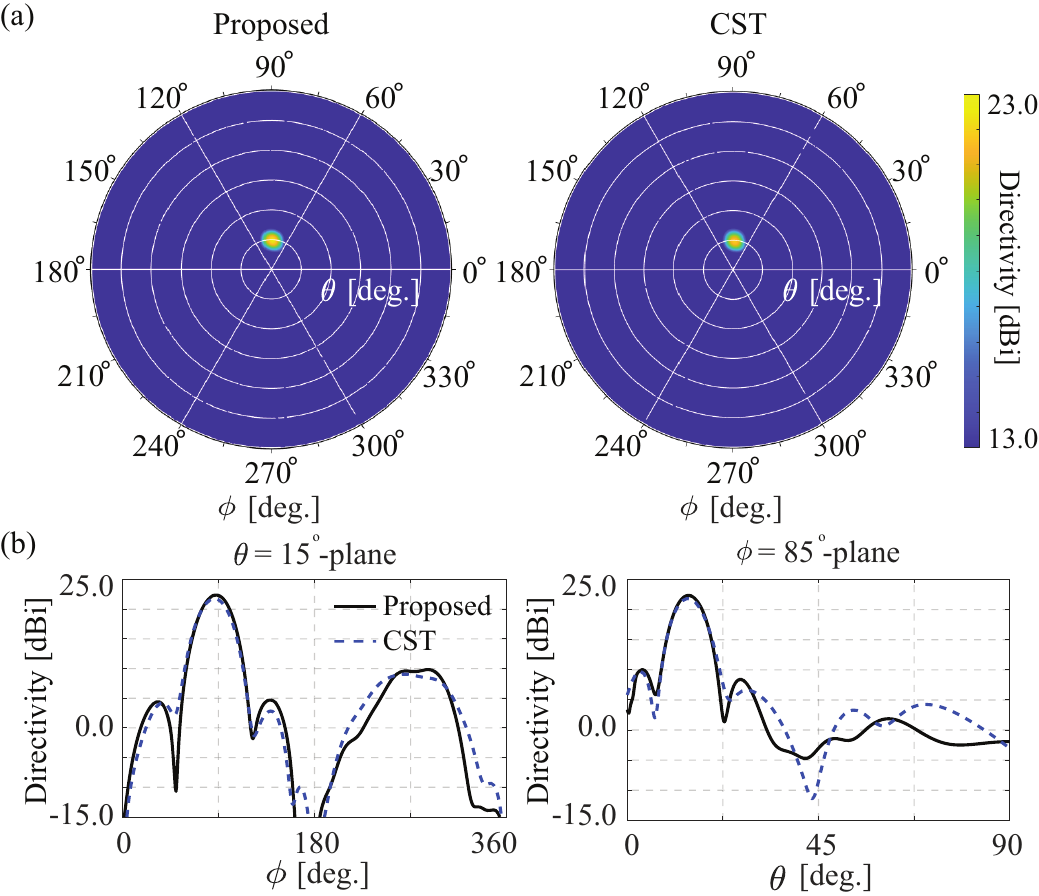}
\caption{(a) Directivity pattern obtained using the proposed model and full-wave analysis. The frequency is $f_{op}=10.0$ GHz. (b) Cross-sectional plots of the directivity patterns at the planes $\theta=15^{\circ}$ (left) and $\phi=85^{\circ}$ (right).}
\label{Fig7_Pattern_Multitile}
\end{figure}


Given the geometry of the multi-tile system, an iterative design procedure is employed using the proposed multi-tile model. The procedure consists of the following steps: solving the coupled linear equations in (\ref{DDA_eq_multitile}) to obtain the dipole moments representing the metamaterial elements and coupling slots, using the initial population; computing the S- and Z-parameters of the metasurface tiles using (\ref{partitioned_matrix}); determining the amplitudes of the voltage waves exciting the metasurface tiles via (\ref{Vinc}), (\ref{I1}), and (\ref{Equivalent_V_I}); solving the coupled linear equations in (\ref{DDA_eq_multitile}) using the updated amplitudes of the voltage waves exciting the metasurface tiles; and calculating the overall radiation patterns and gain to evaluate the predefined cost function. 

The outlined steps are then repeated to update the slot lengths $\mathcal{S}$ using a pattern search optimization algorithm. The cost function used in the design process is defined as
\begin{equation} \label{cost_function2}
\begin{aligned}
\text{cost} = -\Big[G_{dB,multi}\left(f_{op},\mathcal{S},\phi_{tar},\theta_{tar}\right) + SLL_{dB}\left(f_{op},\mathcal{S}\right)\Big],\\
\end{aligned}
\end{equation}
where $G_{dB,multi}$ represents the gain of the overall multi-tile system in the target direction, computed as the product of the directivity and $P_{\mathrm{rad}}/P^{i}$. For this example, the target beam direction was chosen to be $\left(\phi_{tar},\theta_{tar}\right)=\left(85^{\circ},15^{\circ}\right)$.

\begin{figure}[!b]
\centering
\includegraphics[width=3.5in]{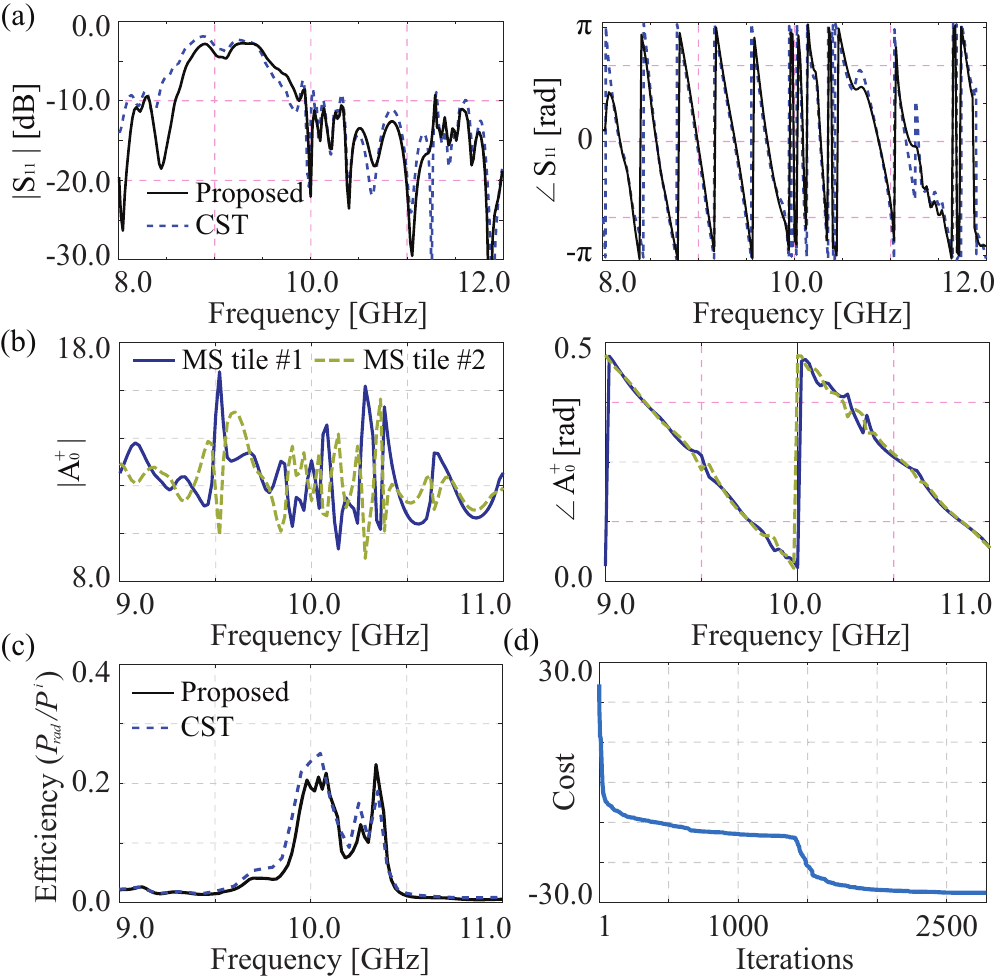}
\caption{(a) The magnitude and phase of $S_{11}$ measured at the input port (i.e., port $1$ of the power divider network) of the multi-tile metasurface system, shown in Fig. \ref{Fig6_NonresonantPolarizability}(a). (b) The amplitudes and phases of the incident voltage waves, computed using the proposed multi-port network analysis technique. (c) The efficiency spectrum of the multi-tile metasurface system, defined as $P_{rad}/P^{i}$, where $P^{i}=0.5$ W. (d) The evolution of the cost function.}
\label{Fig8_NetworkAnalysis_Multitile}
\end{figure}


Figure \ref{Fig7_Pattern_Multitile}(a) compares the directivity pattern at the operating frequency of $f_{op}=10.0$ GHz. In the predicted pattern, the main beam is formed at $\left(\phi^{}_{}=87.8^{\circ}, \theta^{}_{}=14.6^{\circ}\right)$, with a peak value of $22.5$ dB. In the simulated pattern, the main beam reaches a peak value of $21.9$ dB at $\left(\phi^{}_{}=85.7^{\circ}, \theta^{}_{}=14.0^{\circ}\right)$. As such, the radiation pattern predicted using the proposed dipole model closely agrees with that simulated using the full-wave solver, i.e., CST Microwave Studio, confirming the validity of the model. This agreement is further illustrated in the cross-sectional plots of directivity at the target planes, shown in Fig. \ref{Fig7_Pattern_Multitile}(b). It should be noted that the directivity pattern computed using the dipole model accounts for mutual interactions among all radiating slots, as well as those between the metasurface tiles and the power divider.

To better illustrate the validity of the proposed multi-port network analysis technique introduced in Section \ref{multi_port_analysis}, we present key metasurface antenna performance metrics that cannot be readily obtained using conventional approaches, such as the holographic design technique \cite{smith2017analysis} and the cascade model \cite{boyarsky2023cascaded}. Specifically, we compute the magnitude and phase of $S_{11}$ at the input port of the power divider, as shown in Fig. \ref{Fig8_NetworkAnalysis_Multitile}(a). The overall $S_{11}$ of the multi-tile metasurface system predicted by the proposed method is compared with full-wave simulations, demonstrating close agreement over the swept frequencies ranging from $8.0$ to $12.0$ GHz. It is worth noting that the accuracy of the computed $S_{11}$ depends on the S-parameters of the metasurface tiles, modeled as a four-port network, as illustrated in Fig. \ref{Fig2_Schematic_Segmentation}(b). The S-parameters of the metasurface tiles are computed using the proposed coupled dipole model in (\ref{DDA_eq_multitile}) and subsequently incorporated into the multi-port network analysis method (\ref{Zin}) to obtain the overall $S_{11}$ of the multi-tile metasurface system. The S-parameters of the power divider---equivalently, the Z-parameters $\mathbf{Z}^{DIV}$ in Fig. \ref{Fig2_Schematic_Segmentation}(b)---are obtained from full-wave simulations. Therefore, the close agreement between the predicted and simulated S-parameters in Fig. \ref{Fig8_NetworkAnalysis_Multitile}(a) confirms the validity of the proposed design and analysis method. It should be noted that the magnitudes of the predicted and simulated $S_{11}$ exhibit dips of $-22.1$ dB and $-13.8$ dB, respectively, at the operating frequency, indicating that impedance matching is achieved.

We also plot the magnitude and phase of the incident magnetic fields, denoted as $A^{+}_{0}$ in (\ref{Equivalent_V_I}) and (\ref{Hx_inc}), in Fig. \ref{Fig8_NetworkAnalysis_Multitile}(b). As shown in Fig. \ref{Fig8_NetworkAnalysis_Multitile}(b), the magnitudes of the incident fields vary rapidly over the swept frequency range for both metasurface tiles, while the phases remain nearly the same. The equal phases shown in Fig. \ref{Fig8_NetworkAnalysis_Multitile}(b) indicate that the metasurface tiles are excited by waves with the desired equal phase, as specified in Fig. \ref{FigC1_PowerDivider}(b), despite multiple reflections that may occur between the metasurface tiles and the two-way power divider. 

The importance of accurately computing the field amplitudes exciting the metasurfaces (i.e., $A^{+}_{0}$), shown in Fig. \ref{Fig8_NetworkAnalysis_Multitile}(a), can be further demonstrated by analyzing the radiated power of the multi-tile metasurface system. The radiated power of the system in Fig. \ref{Fig6_NonresonantPolarizability}(a) is obtained by computing the far-field pattern and integrating it over the upper hemisphere (i.e., $0 \leq \theta \leq \pi/2$, $0 \leq \phi \leq 2\pi$). Figure \ref{Fig8_NetworkAnalysis_Multitile}(c) compares the ratio of the radiated power, $P_{\mathrm{rad}}$, to the input power at the two-way power divider, $P^{i} = 0.5$ W. To focus on the analysis near the operating frequency of $10.0$ GHz, the efficiency spectrum is shown over $9.0$–$11.0$ GHz. As observed, excellent agreement between the predicted and simulated efficiency spectra (and thus radiated power spectra) is achieved.


It is important to note at this point that accurate computation of the radiated power requires precise evaluation of the far-field pattern, which in turn depends on the accurate determination of the dipole moments representing the metamaterial radiators (i.e., $\mathbf{J}_{ms,1}$ and $\mathbf{J}_{ms,2}$ in (\ref{DDA_eq_multitile}) for this design example). The amplitudes of the dipole moments depend heavily on the incident field amplitudes $A^{+}_{0}$, as they are directly incorporated into the matrices $\mathbf{F}_{ms,1}$ and $\mathbf{F}_{ms,2}$ in (\ref{DDA_eq_multitile}). Therefore, the close agreement shown in Fig. \ref{Fig8_NetworkAnalysis_Multitile}(c) validates the computed excitation fields $A^{+}_{0}$, and consequently the proposed design and analysis approach that combines the coupled dipole model (\ref{DDA_eq_multitile}) with the multi-port network analysis method (\ref{Vinc})-(\ref{Equivalent_V_I}).

We next present the evolution of the cost function defined in (\ref{cost_function}) in Fig. \ref{Fig8_NetworkAnalysis_Multitile}(d) to demonstrate the efficiency of the proposed method. As shown in Fig. \ref{Fig8_NetworkAnalysis_Multitile}(d), the cost function starts at an initial value of $26.6$, exhibits rapid drops in the early stages, and converges to $-27.6$. The optimization, implemented using the pattern search method, was terminated after achieving satisfactory convergence. The rapid convergence, as depicted in Fig. \ref{Fig8_NetworkAnalysis_Multitile}(d), confirms the effectiveness of the proposed design method for the multi-tile metasurface system. It should be noted, however, that the cost function defined in (\ref{cost_function2}) involves the optimization of $288$ tuning parameters (i.e., $\mathcal{S}$), which can become computationally demanding for very large metasurface tile systems. Consequently, advanced optimization approaches that incorporate machine learning techniques \cite{liu2018generative,li2022empowering,ji2023recent,ma2019probabilistic,fromenteze2023morphogenetic} may further improve design efficiency. Note that the primary goal of this work is to demonstrate the validity of the proposed model and its utility as a fast forward-analysis tool, rather than to establish a comprehensive design methodology for the multi-tile metasurface systems.


Lastly, we emphasize that the proposed model for multi-tile metasurface systems achieves excellent agreement with full-wave simulations while significantly reducing computational cost. For the design example shown in Fig. \ref{Fig6_NonresonantPolarizability}(a), the full-wave solver required approximately $1.4$ hours per frequency point on a workstation equipped with a 64-bit 4.20 GHz CPU (24 cores), with a peak memory usage of $345$ GB out of $441$ GB of available memory. This computational cost arose from the discretization of the overall structure into approximately $25.5$ million mesh cells in the frequency-domain solver. Note that, owing to this high computational cost, the overall dimensions of the multi-tile system---including the number of metasurface tiles, constituent waveguides, and metamaterial radiators, as well as their type---were selected to demonstrate improved efficiency near the operating frequency shown in Fig. \ref{Fig8_NetworkAnalysis_Multitile}(c), rather than to present application-specific optimized designs. In contrast, the proposed method, implemented in MATLAB on the same workstation, completes each iteration in approximately $18.2$ seconds per frequency point, including radiated power computation. Such a significant reduction in computation time highlights the efficiency of the proposed approach as a forward model for multi-tile metasurface systems.

\section{Conclusion}
We have presented a semi-analytical model of the multi-tile metasurface antennas and the design using the model. Through numerical simulations, we verified the coupled dipole model for a single tile metasurface, consisting of an arrayed rectangular waveguide-backed metasurfaces and a slotted waveguide feeding the metasurfaces. We then extended the model to multi-tile metasurface systems fed by a practical power divider, and present a design technique that combines the model with a multi-port network analysis to account for electromagnetic interactions between the metasurface tiles and the power divider. The efficacy of the proposed model and design method has been validated via the full-wave numerical analyses, thereby suggesting their applications for the design and optimization of electrically large, multi-tile metasurface systems for next-generation remote sensing and wireless communication systems.

\section*{Appendix A \\Multiport Network Analysis using S-parameters}
\makeatletter
\def\thefigure{A\@arabic\c@figure}
\def\theequation{A\@arabic\c@equation}
\makeatother

\setcounter{equation}{0}
\setcounter{figure}{0}


In this Appendix, we derive the expression for the voltage wave entering the metasurface tile in terms of S-parameters and establish its equivalence to (\ref{Vinc}). This derivation provides an alternative and more intuitive representation of the voltage wave, since S-parameters relate incident and reflected waves, whereas Z-parameters are defined in terms of total voltages and currents.

We begin by defining the S-parameters of the divider, denoted as $\mathbf{S}^{DIV}$, and those of the metasurface tiles terminated with matched loads, denoted as $\mathbf{S}^{MS'}$, as illustrated in Fig. \ref{Fig2_Schematic_Segmentation}(b). The S-parameters are given by
\begin{equation} \label{S_param_def1}
\begin{aligned}
\LaTeXunderbrace{\begin{bmatrix*}[l]
b_{1}^{DIV} \\
b_{2}^{DIV} \\
b_{3}^{DIV}
\end{bmatrix*}}_{\triangleq\mathbf{b}^{DIV}}=
\LaTeXunderbrace{\begin{bmatrix*}[l]
S_{11}^{DIV} & S_{12}^{DIV} & S_{13}^{DIV} \\
S_{21}^{DIV} & S_{22}^{DIV} & S_{23}^{DIV} \\
S_{31}^{DIV} & S_{32}^{DIV} & S_{33}^{DIV} \\
\end{bmatrix*}}_{\triangleq\mathbf{S}^{DIV}}
\LaTeXunderbrace{\begin{bmatrix*}[l]
a_{1}^{DIV} \\
a_{2}^{DIV} \\
a_{3}^{DIV}]^{T}
\end{bmatrix*}}_{\triangleq\mathbf{a}^{DIV}}
\end{aligned}
\end{equation}
and
\begin{equation} \label{S_param_def2}
\begin{aligned}
\LaTeXunderbrace{\begin{bmatrix*}[l]
b_{4}^{MS'} \\
b_{5}^{MS'}
\end{bmatrix*}}_{\triangleq\mathbf{b}^{MS'}}=
\LaTeXunderbrace{\begin{bmatrix*}[l]
S_{44}^{MS'} & S_{45}^{MS'} \\
S_{54}^{MS'} & S_{55}^{MS'} \\
\end{bmatrix*}}_{\triangleq\mathbf{S}^{MS'}}
\LaTeXunderbrace{\begin{bmatrix*}[l]
a_{4}^{MS'} \\
a_{5}^{MS'}
\end{bmatrix*}}_{\triangleq\mathbf{a}^{MS'}}
\end{aligned}
\end{equation}
Note that the interconnection of the two networks results in 
\begin{equation} \label{BC}
\begin{aligned}
\mathbf{a}^{MS'}&=\mathbf{b}^{DIV,r}=[b^{DIV}_{2}, b^{DIV}_{3}]^{T}, \\
\mathbf{b}^{MS'}&=\mathbf{a}^{DIV,r}=[a^{DIV}_{2}, a^{DIV}_{3}]^{T}=\mathbf{S}^{MS'}\mathbf{a}^{MS'}.
\end{aligned}
\end{equation}


Using (\ref{S_param_def1}), the reflected wave vector from the divider ports can be expressed as
\begin{equation} \label{b_div}
\begin{aligned}
\mathbf{b}^{DIV,r}=
\begin{bmatrix}
S_{21}^{DIV} \\
S_{31}^{DIV} \\
\end{bmatrix} a_{1}^{DIV}
+ \mathbf{S}^{DIV,r}\mathbf{a}^{DIV,r},
\end{aligned}
\end{equation}
where
\begin{equation} \label{S_div_r}
\begin{aligned}
\mathbf{S}^{DIV,r}=
\begin{bmatrix}
S_{22}^{DIV} & S_{23}^{DIV} \\
S_{32}^{DIV} & S_{33}^{DIV} \\
\end{bmatrix}.
\end{aligned}
\end{equation}
Substituting (\ref{BC}) into (\ref{b_div}) and rearranging the resulting expression yields
\begin{equation} \label{a_MS_p}
\begin{aligned}
\mathbf{a}^{MS'}=
\mathbf{T}
\begin{bmatrix}
S_{21}^{DIV} \\
S_{31}^{DIV} \\
\end{bmatrix} a_{1}^{DIV},
\end{aligned}
\end{equation}
with $\mathbf{T}$ and $\mathbf{U}$ respectively being the internal coupling term and the identity matrix. The matrix $\mathbf{T}$ is given by
\begin{equation} \label{T_internal}
\begin{aligned}
\mathbf{T}=\left[\mathbf{U}-\mathbf{S}^{DIV,r}\mathbf{S}^{MS'}\right]^{-1}.
\end{aligned}
\end{equation}
Using (\ref{Equivalent_V_I}) and (\ref{a_MS_p}), the net voltage wave entering the metasurface tiles can be expressed as
\begin{equation} \label{V_MS_f}
\begin{aligned}
\mathbf{V}^{+}_{q}=
\mathbf{T}
\begin{bmatrix}
S_{21}^{DIV} \\
S_{31}^{DIV} \\
\end{bmatrix} \sqrt{\frac{2P^{i}}{\text{Re}\{1/Z_{0}\}}},
\end{aligned}
\end{equation}
where $P^{i}$ represents the power incident on the input port of the divider. Note that the standard conversion between S- and Z-parameters, i.e., $\mathbf{S}=\left(\mathbf{Z}-Z_{0}\mathbf{U}\right)^{-1}\left(\mathbf{Z}+Z_{0}\mathbf{U}\right)$ \cite{pozar2009microwave} can be used to convert (\ref{V_MS_f}) into (\ref{Vinc}), with the intermediate steps omitted here for brevity, thereby establishing their equivalence.

\section*{Appendix B \\Design of Slotted Feed Waveguide}
\makeatletter
\def\thefigure{B\@arabic\c@figure}
\def\theequation{B\@arabic\c@equation}
\makeatother

\setcounter{equation}{0}
\setcounter{figure}{0}

\begin{figure}[!b]
\centering
\includegraphics[width=3.5in]{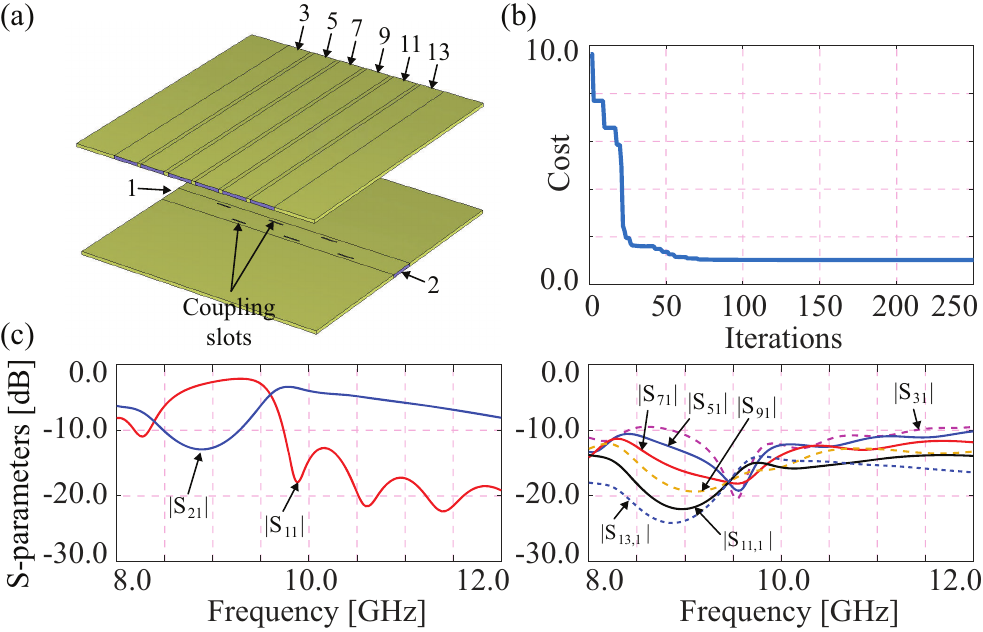}
\caption{(a) Exploded view of an array of rectangular waveguides fed by a slotted waveguide, with a selected set of port numbers indicated. (b) Evolution of the cost function. (c) Selected S-parameters.}
\label{FigB1_FeedWaveguide}
\end{figure}

In this appendix, we present design of a slotted feed waveguide used to excite the metasurface array shown in Fig. \ref{Fig3_ResonantPolarizability}(a). For this purpose, we consider an array of rectangular waveguides fed by a slotted rectangular waveguide, as illustrated in Fig. \ref{FigB1_FeedWaveguide}(a). As shown in Fig. \ref{FigB1_FeedWaveguide}(a), rectilinear coupling slots are inserted into the top wall of the feed waveguide, thereby forming a slotted waveguide structure. The coupling slots are also inserted into the bottom wall of the waveguide array to enable RF power to be coupled for excitation. The coupling slots are positioned at $y=\pm 4.0$ mm to introduce phase diversity in the feed waves, which has been shown to be effective in suppressing the formation of undesired lobes \cite{boyarsky2020grating}.

Given the waveguide geometry, the slot lengths were determined using a reduced version of the dipole model in (\ref{DDA_eq_singletile}) in conjunction with the surrogate optimization method in \cite{queipo2005surrogate}. In this reduced model, only the coupling slots (i.e., $\mathbf{m}_{x}^{S}$) were considered, while the metamaterial elements were omitted. The cost function to be minimized was defined as $\sum_{n=3}^{n=14}|S_{n1}-S_{tar}|$ where $S_{tar}=-15.0$ dB is the target S-parameter. The cost function was formulated to ensure that each waveguide in the array is excited with a comparable level of RF input power while maintaining impedance matching at the input port (i.e., port 1 in Fig. \ref{FigC1_PowerDivider}(a)).


Figure \ref{FigB1_FeedWaveguide}(b) illustrates the evolution of the cost function, demonstrating a rapid convergence to a value of $1.0$. Figure \ref{FigB1_FeedWaveguide}(c) presents the magnitudes of a subset of the S-parameters. As observed, the magnitudes are close to $-15.0$, with a maximum deviation of approximately $\sim 1.12$ dB. Such a level of deviation indicates that the power injected into the input port (i.e., port 1, indicated in Fig. \ref{FigB1_FeedWaveguide}(a)) of the feed waveguide is not perfectly uniformly distributed among the upper waveguides. Although a more uniform power distribution could be achieved by increasing the number of the upper waveguides being fed \cite{yoo2025toward} and/or further optimizing the geometry of the coupling slots \cite{rengarajan1989analysis}, we here aim to validate the proposed analytical model of the multi-tile metasurface configuration shown in Fig. \ref{Fig1_Schematic}, rather than to propose a fully optimized antenna design for specific performance metrics. Accordingly, the investigation of alternative coupling slot configurations is deferred to future work.

\section*{Appendix C \\Design of Power Divider Network}
\makeatletter
\def\thefigure{C\@arabic\c@figure}
\def\theequation{C\@arabic\c@equation}
\makeatother

\setcounter{equation}{0}
\setcounter{figure}{0}

\begin{figure}[!t]
\centering
\includegraphics[width=2.85in]{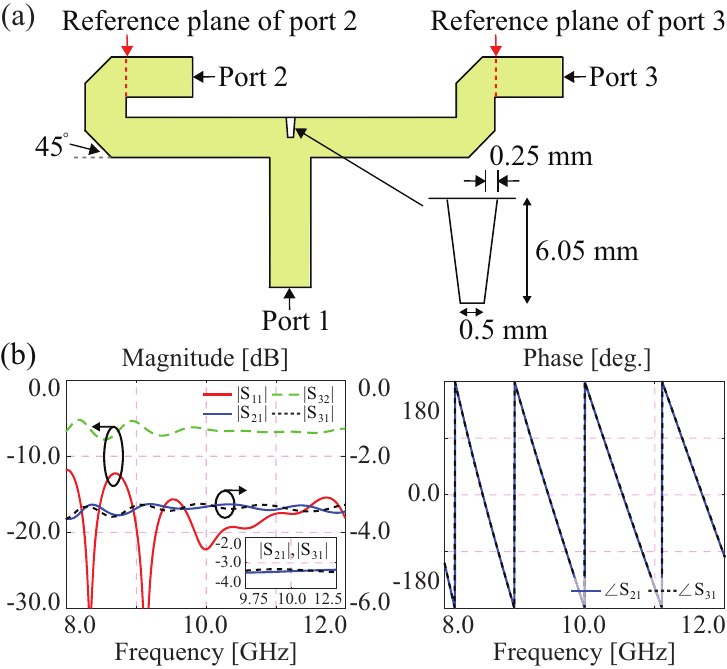}
\caption{(a) Schematic of a two-way power divider network used to excite a multi-tile metasurface with two feed waveguides, shown in Fig. \ref{Fig7_Pattern_Multitile}(a). Port 1 in the divider is the input port, and ports 2 and 3 are the output ports. (b) The magnitude and phase of S-parameters.}
\label{FigC1_PowerDivider}
\end{figure}

In this appendix, we present design of a two-way power divider network used to feed the two metasurface tiles shown in Fig.~\ref{Fig6_NonresonantPolarizability}(a). It should be emphasized that the aim of this appendix is to provide a divider design that supports the proposed multi-tile metasurface design methodology, rather than to develop a practical divider highly optimized for specific applications.

The schematic of the designed power divider is shown in Fig. \ref{FigC1_PowerDivider}(a). The divider consists of rectangular waveguide sections with a trapezoidal wedge and mitered bends. The symmetric trapezoidal wedge is located at the center of the structure for equal power division of RF signal injected into the input port (i.e., port $1$), while the mitered bends are used to route the signals to the output ports (i.e., ports $2$ and $3$) with reduced return and insertion losses. The reference planes for the output ports, indicated in Fig. \ref{FigC1_PowerDivider}(a), are defined at the input ports of the corresponding metasurface tiles. For accurate characterization, additional waveguide sections were added to mitigate the influence of higher-order modes generated by the mitered bends on the computed S-parameters. The deembedding technique was subsequently applied to shift the reference planes to match those of the divider used in Fig. \ref{FigC1_PowerDivider}(a). 

The simulated S-parameters are shown in Fig. \ref{FigC1_PowerDivider}(b), confirming the intended operation of the divider. Specifically, the input RF signal is evenly divided and delivered to the output ports with equal magnitude (i.e., $|S_{21}|=-3.3$ dB, $|S_{31}|=-3.3$ dB) and phase (i.e., $\angle S_{21}=-179.9^{\circ}$, $\angle S_{31}=179.9^{\circ}$), while acheiving the impedance matching (i.e., $|S_{11}|<-20.0$ dB), at the target frequency $f_{op}=10.0$.


\section*{Acknowledgment}
This material is based upon work supported by the Defense Advanced Research Projects Agency (DARPA) and Naval Information Warfare Center Pacific (NIWC Pacific) under Contract No. N66001-21-C-4016. This work was also supported by the National Research Foundation of Korea (NRF) grant funded by the Korea government (MSIT, Grant No.: RS-2025-00554900) and by the Yonsei University Research Fund (Grant No.: 2024-22-0062).

\bibliographystyle{IEEEtran}
\bibliography{references.bib}

\end{document}